\documentclass[%
 reprint,superscriptaddress,
 amsmath,amssymb,
 aps,
pre,
]{revtex4-2}
\usepackage{lipsum}
\usepackage{hhline}
\usepackage{graphicx}
\usepackage{dcolumn}
\usepackage{bm}
\usepackage{hyperref}
\usepackage{color}
\usepackage{subfigure}
\usepackage[normalem]{ulem}

\begin{document}

\preprint{}

\title{\texorpdfstring{Unconventional criticality, scaling breakdown, and diverse universality classes \\in the Wilson-Cowan model of neural dynamics}{TEXT}}

\author{Helena Christina Piuvezam}
    \email{hcpiuvezam@gmail.com}
 \affiliation{Departamento de F\'{\i}sica, Universidade Federal de Pernambuco, Recife PE 50670-901, Brazil.}
\author{Bóris Marin}%
\affiliation{Centro de Matemática, Computação e Cognição, Universidade Federal do ABC, São Bernardo do Campo, Brazil}
\author{Mauro Copelli}%
\affiliation{Departamento de F\'{\i}sica, Universidade Federal de Pernambuco, Recife PE 50670-901, Brazil.}
\author{Miguel A. Muñoz}%
 \email{mamunoz@onsager.ugr.es}
\affiliation{Instituto Carlos I de Física Teórica y Computacional, Universidad de Granada, Granada, Spain.}

\date{\today}

\begin{abstract}
The Wilson-Cowan model constitutes a paradigmatic approach to understanding the collective dynamics of networks of excitatory and inhibitory units. It has been profusely used in the literature to analyze the possible phases of neural networks at a mean-field level, e.g., assuming large fully-connected networks. Moreover, its stochastic counterpart allows one to study fluctuation-induced phenomena, such as avalanches. 
Here, we revisit the stochastic Wilson-Cowan model paying special attention to the possible phase transitions between quiescent and active phases. 
We unveil eight possible types of phase transitions, including continuous ones with scaling behavior belonging to known universality classes ---such as directed percolation and tricritical directed percolation--- as well as novel ones. 
In particular, we show that under some special circumstances, at a so-called Hopf tricritical directed percolation transition, rather unconventional behavior  including an anomalous breakdown of scaling
emerges.
These results broaden our knowledge of the possible types of critical behavior in networks of excitatory and inhibitory units and are of relevance to understanding avalanche dynamics in actual neuronal recordings.  From a more general perspective, these results help extend the theory of non-equilibrium phase transitions into quiescent or absorbing states.
\end{abstract}


\maketitle


\section{Introduction}\label{sec:intro}

A large variety of natural systems exhibit continuous (second-order) phase transitions between an active phase and a quiescent (or absorbing) one where all activity ceases~\citep{Marro, Hinrichsen, GGMA, Henkel, Odor}. 
These systems often exhibit scaling behavior around the phase-transition point and this is typically described by the \emph{directed percolation} universality class, as originally conjectured by Janssen and Grassberger~\citep{Janssen, Grassberger}.
Actually, directed percolation (DP) is one of the most robust classes of universal critical behavior away from thermal equilibrium~\citep{Marro, Hinrichsen, GGMA, Henkel, Odor, Binney}, as it describes all possible phase transitions into an absorbing state ---even for multi-component systems~\citep{GG}--- in the absence of additional symmetries or conservation laws~\citep{Henkel, GGMA, Marro, Odor,many}.
Moreover, some of the representative models of this class, such as the branching process and the contact process~\citep{Harris, Liggett}, have been broadly studied in a large variety of contexts, including countless applications in materials science, turbulence, epidemics, theoretical ecology, social sciences, and neuroscience.

Conversely, under some circumstances, phase transitions into quiescent states occur in a discontinuous (or first-order) rather than continuous manner.
This is often the case when higher-order reactions are considered, where at least a pair of active units are required to activate the third one~\citep{Lubeck06, Odor, Villa}.
This situation usually involves a bistable regime (i.e., with phase coexistence), leading to hysteresis.
There are also well-studied systems (see e.g., a modified contact process~\cite{Lubeck06, Assis09}) that include both types of transitions, continuous and discontinuous, as well as a \emph{tricritical} point with a scaling behavior that differs from DP and is described by the so-called tricritical directed percolation (TDP) universality class~\citep{Lubeck06}.

In the context of neuronal systems, the experimental work by Beggs and Plenz reported on the existence of \emph{neuronal avalanches} (i.e., outbursts of neuronal activity between quiescent periods). 
These exhibited highly-variable sizes and durations, which were power-law distributed.
Moreover, the associated exponents were found to be consistent with those of critical systems in the mean-field DP universality class~\citep{BP2003}, suggesting that brain dynamics could be poised near the edge of a phase transition~\citep{Mora, Chialvo2010, RMP,Plenz-review,Chialvo-review,Obyrne}. 
Further experimental works reported evidence of some scaling exponents that deviate from DP~\citep{Fontenele19, zebra}, so that the interpretation of the scaling behavior in terms of universality classes remains a current matter of debate~\citep{RMP}.
In particular, the possible departure from the standard DP class (together with the possible existence of discontinuous transitions in brain dynamics \citep{Millman,neutral,Cortes}) raises a number of questions from the theoretical point of view.
For example, the fact that neuronal networks include inhibitory units --- which hinder activity propagation and are not usually included in simple models in the DP class, such as the standard branching process --- triggered a renewed interest in the scrutiny of alternative types of critical behavior (as well as discontinuous transitions and tricriticality) in networks of excitatory and inhibitory units~\citep{Benayoun, Assis09,Kinouchi, Girardi,girardi0,tawan21,Lucilla,deCandia21,nandi22,Apicella,Corral22}.
Do different types of quiescent to active phase transitions emerge in simple models of activity propagation once inhibitory effects are considered?

Here, to further advance our knowledge along these lines, we study one of the most broadly studied parsimonious models in neuroscience: the Wilson-Cowan model~\citep{WilsonCowan72} as well as its stochastic counterpart~\citep{Benayoun,Corral22}. 
We systematically analyze the resulting phase diagram, the possible phases and phase transitions.
In particular, we reveal that, depending on the relative strengths of excitatory and inhibitory couplings, there can be up to $8$ different types of phase transitions into quiescence.
Some of them exhibit well-known scaling behavior (such as DP or TDP), while others are discontinuous or show different types of anomalies in scaling or even mixed features of continuous and discontinuous transitions.
Finally, we elucidate a novel type of phase transition that is highly non-trivial, exhibiting unconventional behavior and breakdown of scaling.

Our results help rationalize and categorize the possible types of criticality in networks of excitatory and inhibitory units, contributing to the advance of the brain-criticality hypothesis and of the general theory of non-equilibrium phase transitions~\citep{Henkel}.

\begin{figure}[t]
\centering
{{\includegraphics[width = 0.85\columnwidth]{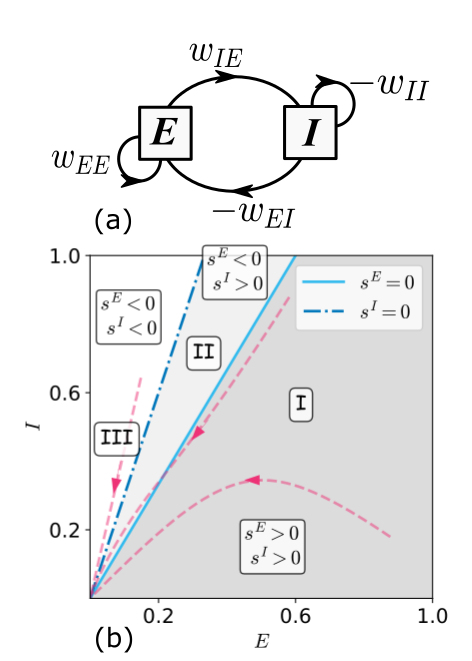}}}
\caption{(a) Sketch of the Wilson-Cowan model, including both excitatory and inhibitory populations
and synaptic couplings between them.
(b) The piecewise-smooth nature of the response function $\Phi$ in Eq.~\eqref{eq:response_function} generates three different regions in the state space, depending on the sign of the total input $s$: in region I, $s^E$ and $s^I$ are both positive; in region II, $s^E$ is negative and $s^I$ is positive; and, finally, in region III, $s^E$ and $s^I$ are negative (the figure illustrates these regions
for $w_{EE}=2.5$, $w_{EI}=1.5$, $w_{IE}=1.5$, $w_{II}=0.5$, and $h = 0$).
Observe that when $w_{EE}/w_{EI} > w_{IE}/w_{II}$, lines $s^E=0$ and $s^I=0$ switch position and region II then shows $s^E > 0$ and $s^I < 0$. 
This condition is not explicitly explored because the analytical results are preserved under it.
The trajectories, shown as red dashed arrows, illustrate how the system is attracted to the origin
(quiescent state). 
Trajectories that start from initial conditions $E(0) > I(0)$ typically decay to zero through region I. 
As discussed in Section~\ref{sec:scaling-properties},
initial conditions in region I that are close to the \emph{switching manifold} $s^{E}=0$ can cross over to region II.
Furthermore, regions II and III are trapping:  
once trajectories cross the switching manifold, they are unable to return.
}
\label{fig:regions}
\end{figure}

\section{The Wilson-Cowan Model and its stochastic counterpart}

In its original formulation, the Wilson-Cowan model describes the collective deterministic (or ``mean-field'') behavior of a local population of both excitatory and inhibitory neurons by means of two coupled differential equations~\citep{WilsonCowan72}. 
These equations reproduce ---as a function of a set of coupling-strength parameters--- a variety of possible dynamical regimes, all of which with counterparts in actual neuronal systems~\citep{Wallace11, deCandia21, Maruyama14, WC-review} that are delimited by phase transitions (bifurcation lines)~\citep{hoppensteadt1997weakly, Borisyuk}. 

To go beyond this deterministic or mean-field picture, Benayoun {\it et al.}~\citep{Benayoun} proposed a microscopic version of the Wilson-Cowan model in the form of a Markovian process for a population of coupled excitatory and inhibitory individual binary neurons that can be either active or inactive~\footnote{Let us remark  that a very similar model has been recently proposed to generalize the standard contact process to include inhibitory units, and is described by similar equations~\citep{Corral22}).}. 
In the \emph{stochastic Wilson-Cowan} (SWC) model, the state of each unit  $\ell$ at a given time $t$--- which can be either excitatory (E) or inhibitory (I) --- is given by $\sigma_\ell^{E/I}(t) = 1$ for active neurons and $\sigma_\ell^{E/I}(t)=0$ for inactive ones.
These state variables change according to a master equation specified by transition rates defined as follows.

Each active neuron, regardless of its type, shifts from the active to the quiescent state (decay), $1 \rightarrow 0$, at a constant rate $\alpha$. 
The reverse transition (activation), $0 \rightarrow 1$, occurs at rate $\Phi(s^{\ell})$, defined as
\begin{eqnarray}
\Phi(s^{\ell}) &=& \left\{
\begin{array}{l l}
\tanh(s^{\ell}), & \text{if } s^{\ell}>0\\
0, & \text{otherwise,}
\end{array}
\right.\label{eq:response_function}
\end{eqnarray}
where the \emph{input} $s^{\ell}$ to neuron, ${\ell}$ is
\begin{eqnarray}
s^{\ell} &=& \sum_m w^{\ell m}\sigma_m + h,
\end{eqnarray}
$w^{\ell m}$ is the synaptic weight from neuron $m$ to neuron $\ell$, and $h$ is a constant external input. 
Observe that the form of the \emph{response function}, $\Phi(s^\ell)$, in Eq.~\eqref{eq:response_function} enforces the non-negativity of the transition rates.
In what follows, the synaptic weights are chosen to depend only on the type (excitatory or inhibitory) of both the pre-synaptic and the post-synaptic neuron, leaving (as sketched in Fig.~\ref{fig:regions}a) only four free parameters: $w^{\ell m} \in \{w_{EE}, w_{IE}, w_{IE}, w_{II} \} ~~\forall \ell, m$, where, e.g., $w_{IE}$ is the excitatory coupling strength to inhibitory neurons, and so forth. Previous works on this model have often employed symmetric weights as a way to reduce the dimensionality of the phase diagram; e.g., common excitatory ($w_{E} \equiv w_{EE} = w_{IE}$) and inhibitory ($w_I \equiv w_{II} = w_{EI}$) inputs~\citep{Brunel00, Benayoun, deCandia21}.
In order to systematically explore the full set of possible phase transitions, here, we remove such constraints.

This stochastic process can be implemented on different types of networks, as specified by a \emph{connectivity matrix}. 
As a first approach, one can assume a large fully-connected network of size $N$. 
Indeed, performing a (network) size expansion~\citep{vankampen2007spp, gardiner2004handbook}, one  recovers ---up to leading-order--- the standard Wilson-Cowan equations~\citep{Benayoun}, written as:
\begin{eqnarray}
\dot{E} &=& - \alpha E + \left(1-E\right)\Phi\left(w_{EE}E-w_{EI}I+h\right),\label{eq:WCEI-Eeq}\\
\dot{I}\; &=& -\alpha I + \left(1\;-\;I\right)\Phi\left(w_{IE}E - w_{II}\;I + h\right)\; ,\label{eq:WCEI-Ieq}
\end{eqnarray}
where $E$ and $I$ are the densities of active excitatory and inhibitory neurons, respectively~\citep{Benayoun} and $\alpha$ is the decay rate.
Similarly, by adding next-to-leading corrections, one obtains a set of two Langevin equations including square-root noise (similar to the ones for DP and TDP), which we do not write explicitly here.
These stochastic equations allow one to describe fluctuation effects in finite-size (fully connected) networks~\citep{vankampen2007spp, ohira1997, Benayoun} ---though the forthcoming computational analyses refer to simulations of the microscopic model--- as well as to perform a systematic scaling analysis of the full model.

Observe that, owing to the discontinuous derivative of $\Phi$ at zero, Eq.~\eqref{eq:WCEI-Eeq} and Eq.~\eqref{eq:WCEI-Ieq} are piecewise smooth differential equations~\citep{Piecewise19, Piecewise_WC, Kunze} --- i.e., they are smooth everywhere except at \emph{switching manifolds}, which are defined by the conditions of vanishing input in the response function: $s^{E} \equiv w_{EE}E-w_{EI}I+h = 0$ and $s^{I} \equiv w_{IE}E - w_{II}\;I + h = 0$.
These two conditions divide the state space into three regions: I, II, and III, as illustrated in Fig.~\ref{fig:regions}b:
\begin{itemize}
    \item In region II, Eq.~\eqref{eq:WCEI-Eeq} becomes $\dot{E} = -\alpha E$ and trajectories in this region decay exponentially fast to the quiescent phase (either crossing to region III or not).
    \item Similarly, in region III,  $\dot{E} = -\alpha E$ and also $\dot{I} = -\alpha I$, leading to an even faster decay to quiescence.
    \item Conversely, in region I, the total input does not vanish for either sub-population and the dynamics can be more complex, possibly reaching non-trivial (active) fixed points.
\end{itemize}
Inspection of Eq.~\eqref{eq:WCEI-Eeq} and Eq.~\eqref{eq:WCEI-Ieq} readily reveals that trajectories starting in regions II and III do not cross over to region I (as excitation always diminishes in these regimes), but the opposite can  happen (see, e.g., the central trajectory shown in Fig.1 as well as Appendix~\ref{ap:trap}).

In the next sections, we explore in detail, both analytically and numerically, the features of each of the possible phase diagrams as well as all the possible  phase transitions between quiescent and active states.

\section{\texorpdfstring{Mean-field Phase Diagrams: \\General and specific features}{TEXT}}

To avoid confusion, let us first underline that in what follows we refer indistinctly to \emph{phase transitions} or to \emph{bifurcations}, as the present  focus is on the description of fully-connected networks (i.e., mean-field systems).
Thereby, DP transitions correspond to transcritical bifurcations, discontinuous transitions to saddle-node bifurcations, tricritical points to saddle-node-transcritical (codimension-2) bifurcations~\citep{Izhi,Strogatz}, and so on.

In the absence of any external driving force ($h = 0$), the steady-state conditions for Eq.~\eqref{eq:WCEI-Eeq} and~ Eq.\eqref{eq:WCEI-Ieq} always admit a trivial solution $E^*=I^*=0$, which defines the \emph{quiescent phase} as well as, possibly, some 
non-trivial solutions ($E^*> 0$ and $I^*>0$) of the following equations,
\begin{eqnarray}
E^* &=& \frac{1}{w_{IE}}\left[w_{II} I^* + \Phi^{-1}\left(\frac{\alpha I^*}{1 - I^*}\right)\right], \label{eq:EI_fixed_point}\\
I^* &=& \frac{1}{w_{EI}}\left[w_{EE}E^* - \Phi^{-1}\left(\frac{\alpha E^*}{1 - E^*}\right)\right]\,
\label{eq:IE_fixed_point}
\end{eqnarray}
and define the \emph{active phase}.
Observe that Eq.~\eqref{eq:EI_fixed_point} and Eq.~\eqref{eq:IE_fixed_point} are well-defined only as long as $\Phi^{-1}$ exists, i.e., in region~I (Fig.~\ref{fig:regions}b), so that non-trivial solutions exist only inside said region.

Let us now analyze the overall phase diagram, describing the stable phases as a function of the model parameters. 
In particular, without loss of generality, we keep the activity-decay rate $\alpha \neq 0$ and the self-inhibition weight $w_{II}\geq 0$ fixed.
Choosing $w_{EE}$ and $w_{EI}$ as control parameters, depending on the value of the remaining free parameter, $w_{IE}$, the system may display three qualitatively different types of phase diagrams in the $(w_{EE}, w_{EI})$ plane. 
Other parameter choices are possible, but the system is always described by one of these three qualitatively different types of phase diagrams.

\subsection{Quiescent phase and its stability limits}

First of all, let us stress that the quiescent phase is always stable (and is the only stable state) with respect to the introduction of \emph{inhibition-dominated} perturbations, i.e., in regions II and III, so in what follows we focus on its stability and the resulting phase diagram as a result of \emph{excitation-dominated} perturbations.

\begin{figure}[t]
\centering
{\includegraphics[width = \columnwidth]{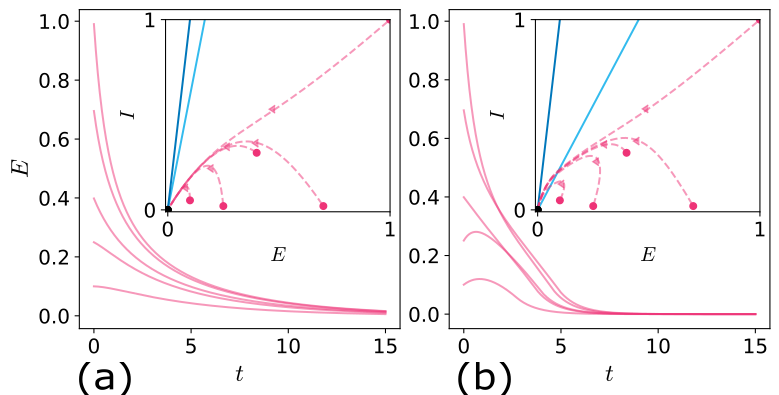}}
\caption{Two different types of quiescent phases. Time series towards the absorbing state of trajectories (a) on the standard quiescent phase ($w_{EE} = 1.2$ and $w_{IE} = 0.2$) and (b) on the excitable quiescent phase ($w_{EE} = 2.2$ and $w_{IE} = 1.0$). 
Observe the non-monotonicity in the second case, which is a manifestation of the excitability of the quiescent state: perturbations can be  amplified before trajectories finally decay to quiescence.
The insets show the phase space for these two cases,  respectively, as well as the corresponding switching manifolds and some sample trajectories (arrows). 
In the second case, trajectories cross the switching manifold.
Parameter values  are $\alpha = 1.0$, $w_{II} = 0.2$, and $w_{EI} = 2.0$.}
\label{fig:quiescent}
\end{figure}

\begin{figure*}[!t]
\centering
\includegraphics[width = \textwidth]{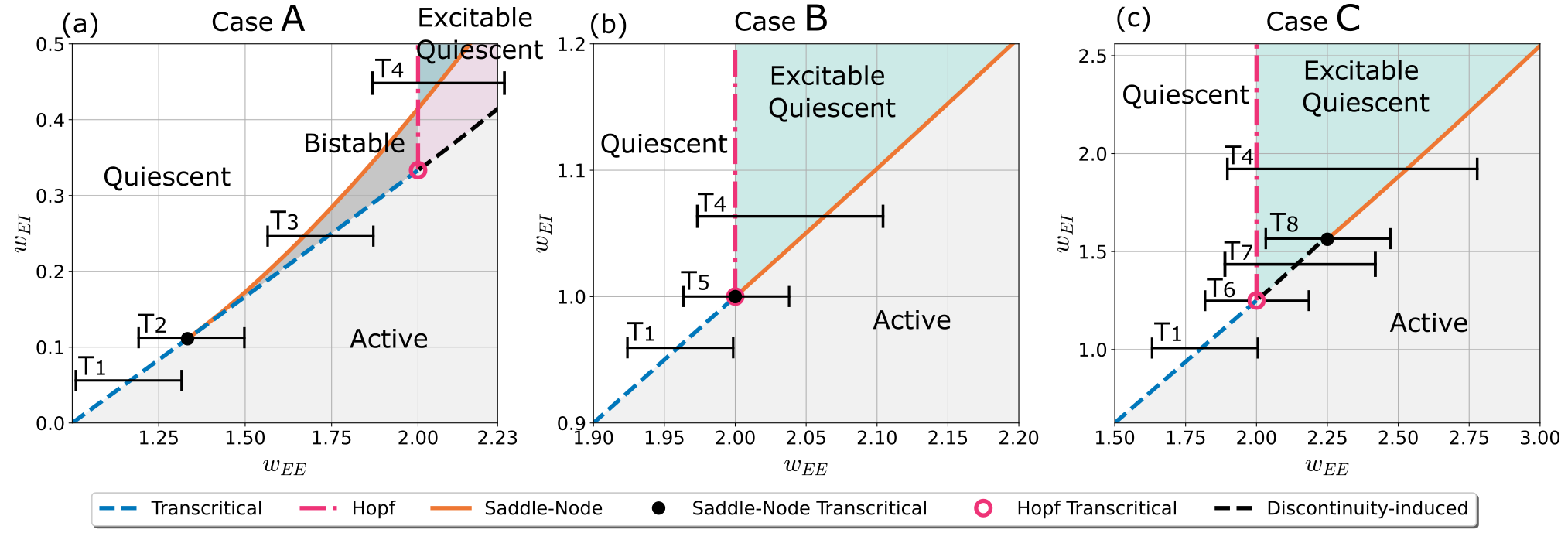}
\caption{Different phase diagrams of the model (Eq.~\eqref{eq:WCEI-Eeq} and Eq.~\eqref{eq:WCEI-Ieq}), with excitation-dominated initial conditions and parameter values: $\alpha = 1$, $w_{II} = 0$, and $w_{IE}=3$ (Case A), $w_{IE}=1$ (Case B), and $w_{IE}=0.8$ (Case C). 
In all cases, the (vertical dot-dashed) line of Hopf bifurcations separates the standard quiescent phase from the excitable quiescent one.
(a) In case A, the Hopf line collides with the (diagonal and dashed) transcritical line to the right of the tricritical point (black dot), within a region of bistability.
(c) In case C, the situation is reversed and the intersection of the Hopf line with the transcritical line occurs to the left of the tricritical point, defining a Hopf-transcritical  bifurcation.
Observe that a line of bifurcations, induced by the piecewise smooth nature of the system, unfolds from the Hopf-transcritical point (dashed black line).
(b) Between the previous two cases, case B (for which fine-tuning a third parameter, $w_{IE}=1$, is needed) the Hopf line collides with the transcritical one at a codimension 3 bifurcation point that we call Hopf-tricritical point (Hopf saddle-node-transcritical bifurcation). 
The horizontal black segments $T_1, \ldots, T_8$ represent $8$ qualitatively different ways to transition from a quiescent to an active state as the control parameter, $w_{EE}$, increases.}
\label{fig:phase-diagram}
\end{figure*}

Importantly, there are two different types of quiescent phases:
(i) The first one is a standard quiescent one, i.e., a regime in which the quiescent phase is locally stable to excitation-dominated perturbations (Fig.~\ref{fig:quiescent}a). 
This occurs if the eigenvalues of the associated stability matrix, as specified by:
\begin{eqnarray}
\lambda_\pm &=& \frac{w_{EE} - 2\alpha - w_{II} \pm \sqrt{(w_{EE} + w_{II})^2 - 4w_{EI}w_{IE}}}{2} \; ,\nonumber\\
\label{eq:eigenvalues}
\end{eqnarray}
have negative real parts (white zone in the diagrams of Fig.~\ref{fig:phase-diagram}).

(ii) Alternatively, if the eigenvalues have positive real parts and an imaginary component, then, in principle, one could expect oscillations away from quiescence to emerge.
However, given the non-smooth piecewise dynamics, the resulting ``curvy'' trajectories end up crossing over to region II, where the dynamics follow the equation $\dot{E} = -\alpha E$ and the quiescent phase is the only attractor.
Therefore, in this regime, small excitatory perturbations to the quiescent phase may give rise to large trajectories in state space before returning to quiescence (see Fig.~\ref{fig:quiescent} and~\citep{Benayoun,Corral22}). This property is called "excitability" (or ``reactivity"~\citep{Serena-NN}) and the corresponding quiescent state is called "excitable quiescent". 

In both cases, either when the quiescent state is standard or excitable, it loses its stability when the real part of the largest eigenvalue becomes positive, which (from Eq.(\ref{eq:eigenvalues})) occurs at
\begin{equation}
w_{EE}^T = \alpha + \frac{w_{EI}w_{IE}}{\alpha + w_{II}}
\;.
\label{eq:transc_wEE}
\end{equation} 
Not surprisingly, separating the previous two phases (standard quiescent and excitable quiescent), there is a line of (supercritical) Hopf bifurcations (dot-dashed vertical lines in Fig.~\ref{fig:phase-diagram}) at
\begin{equation}
\label{eq:AH_wEE}
w^{H}_{EE} = 2\alpha + w_{II}\; 
\end{equation}
with the additional constraint that there is a non-vanishing imaginary part, i.e., from Eq.\eqref{eq:eigenvalues}:
\begin{equation}
(w_{EE} + w_{II})^2 - 4w_{EI}w_{IE} < 0
\end{equation}
(so that the bifurcation is only defined above the Hopf-transcritical line).

Summing up, there are two types of quiescent phases, separated by a line of Hopf bifurcations:
\begin{itemize}
    \item A \textbf{standard quiescent phase}, which is a locally stable node (upper left regions in Fig.~\ref{fig:phase-diagram});
    \item An \textbf{excitable quiescent phase}, which is a locally stable focus (upper right regions in Fig.~\ref{fig:phase-diagram}).
\end{itemize}

\subsection{Active phase and its stability limits}

The active phase becomes a stable solution either at (i) a transcritical bifurcation (i.e., it emerges continuously once the quiescent phase loses its stability in a DP transition), which occurs for Eq.~(\ref{eq:transc_wEE}) as represented by the dashed lines in Fig.~\ref{fig:phase-diagram};
(ii) a saddle-node bifurcation, i.e, emerging discontinuously (solid line in Fig.~\ref{fig:phase-diagram}) at
\begin{equation}
w^{SN}_{EE} = \min_{E^*} \left[\frac{w_{EI}I^*(E^*)}{E^*} + \frac{1}{E^*}\Phi^{-1}\left(\frac{\alpha E^*}{1 - E^*}\right)\right]  \; , 
\label{eq:w_EE_function}
\end{equation}
where $E^*$ and $I^*$ are solutions of Eq.~\eqref{eq:EI_fixed_point} and Eq.\eqref{eq:IE_fixed_point} that can be solved numerically; 
or (iii) at a tricritical point, where the previous two lines meet (black dot in Fig.~\ref{fig:phase-diagram}), to which one can also refer as ``saddle-node-transcritical" (SNT) point (its location $(w^{SNT}_{EE}, w^{SNT}_{EI})$ in the phase diagram is explicitly derived in Appendix~\ref{ap:mat_TDP}; see, in particular, Eq.\eqref{eq:SNT_wei_A} and Eq.~\eqref{eq:SNT_wee_A}).

\subsection{Relative location of the line of  Hopf bifurcations}

Observe that the line of Hopf bifurcations --- which as shown in Fig.~\ref{fig:phase-diagram}a is always vertical in the $(w_{EE}, w_{EI})$ plane --- collides with the line of transcritical bifurcations at a special point (here named Hopf-transcritical (HT) bifurcation, which is marked with an empty circle in the different panels of Fig.~\ref{fig:phase-diagram}).
From Eq.(\ref{eq:AH_wEE}) and Eq.(\ref{eq:transc_wEE}) one can easily derive the conditions for the HT point to occur:
\begin{eqnarray}
w^{HT}_{EI} &=& \frac{(\alpha + w_{II})^2}{w_{IE}},
\label{eq:AHT_wei}\\
w^{HT}_{EE} &=& 2 \alpha + w_{II}.
\label{eq:AHT_wee}
\end{eqnarray}

The key aspect distinguishing the three possible topologies of the phase diagram is whether this HT point lies to the right (case A), left (case C), or on top of the tricritical (SNT) point (case B) in phase space, i.e.:
\begin{itemize}
    \item Case A: $w^{SNT}_{EE} < w^{HT}_{EE}$ 
    \item Case B: $w^{SNT}_{EE} = w^{HT}_{EE}$ 
    \item Case C: $w^{SNT}_{EE} > w^{HT}_{EE}$. 
\end{itemize}
As already mentioned, these three possibilities are illustrated in Fig.~\ref{fig:phase-diagram}, in which the value of $w_{IE}$ changes to switch from one regime to the other. 
Also, note that case B requires a higher level of fine-tuning than the other two cases, which appear in broad regions of parameter space.
From here on, one needs to separately discuss the three aforementioned possible structures of the phase diagram.

\subsubsection{Case A: Left panel in Fig.~\ref{fig:phase-diagram}}

In this case, the HT point lies to the right of the tricritical point.
Visual inspection of Fig.~\ref{fig:phase-diagram}A reveals  that there are four different ways to go from a quiescent phase (either standard or excitable) to the active one. These are labeled as: $T_1$, for the transcritical bifurcation from the standard quiescent, $T_2$, for a standard tricritical transition; $T_3$, for a transition through a bistable regime (saddle-node bifurcation with coexistence between the standard quiescent and the active phase), and $T_4$, also for a discontinuous transition with bistability, although in this case,  between the excitable quiescent phase and the active one.

\subsubsection{Case B: Central panel in Fig.~\ref{fig:phase-diagram}}

Here, the HT point lies exactly on top of the tricritical point.
This structure lies to only three possible types of transitions: $T_1$ and $T_4$ (as already described), and a new transition labeled $T_5$, which occurs through the tricritical (SNT) point that coincides with the special HT point in a codimension 3 bifurcation.

\subsubsection{Case C: Right panel in Fig.~\ref{fig:phase-diagram}}

In this last case, the HT point lies to the left of the tricritical point and there are five types of transitions, including the standard transcritical ($T_1$)  and saddle-node ($T_4$) ones,  as well as three novel ones: $T_6$, a transition through the special HT point; $T_7$, a transcritical bifurcation but into the excitable quiescent phase; and, finally, $T_8$, a tricritical (or SNT) transition into the excitable quiescent phase.

In the next section, we analyze these eight types of phase transitions (or bifurcations) ---from $T_1$ to $T_8$--- scrutinizing the corresponding peculiarities for each of them.

\section{Scaling properties at the different types of transitions\label{sec:scaling-properties}}

Standard linear stability analysis of the fixed points of the (mean-field) dynamics, Eq.~\eqref{eq:WCEI-Eeq} and Eq.~\eqref{eq:WCEI-Ieq}, allows one to study the nature of bifurcations and make analytical predictions for the scaling behavior~\citep{Marro,Henkel,munoz1999}. 
In particular, a linear approximation of Eq.~\eqref{eq:EI_fixed_point} around the quiescent solution yields a value of $I^*$ proportional to the density of active excitatory neurons $E^*$, hence in what follows we employ indistinctly either the latter or the sum of both as an order parameter.

In all cases and for all possible types of transitions, we compute the usual quantities and scaling exponents customarily employed in the analysis of quiescent-active phase transitions (as long as they are well-defined).
Even if, generally, three independent exponent values suffice to fully determine the universality class~\citep{marro_dickman_1999, munoz1999}, here, for the sake of completeness, we compute more, which also allows us to check for consistency. 
We compute ``\emph{static exponents}": such as {\bf (i)} $\beta$, the control parameter one ($E^* \propto \Delta^\beta$), where $\Delta = w_{EE}-w^*_{EE}$ is the distance to the transition point and $w^*_{EE}$ stands, generically, for the value of $w_{EE}$ at the specified bifurcation; 
{\bf (ii)} $\delta_h$, defined by $E^* \propto h^{1/\delta_h}$, representing the response to a constant external field $h$, at criticality. 
"\emph{Correlation exponents}" ($\nu$) such as the
one for {\bf (iii)} the correlation length, $\xi_\perp$,~ 
$\xi_\perp\propto \Delta^{\nu_\perp}$ 
and for {\bf (iv)} the correlation time, $\xi_\parallel$, ~$\xi_\parallel \propto \Delta^{\nu_\parallel}$. 
 "\emph{Dynamic exponents}": such as {\bf (v)} $\theta$, that governs the time decay of the order parameter $E(t)\propto t^{-\theta}$. 
``\emph{Spreading exponents}" such as those describing:
{\bf (vi)} the total number of active sites, $N(t)\propto t^{\eta}$;  
 {\bf (vii)} the mean-squared radius in surviving runs $R^2(t)\propto t^{-z}$; 
and {\bf (viii)} the survival probability $P_s(t)\propto t^{-\delta}$~\citep{Marro}, 
as well as ``\emph{avalanche exponents}"  defined by: 
{\bf (ix)} $P(S)\sim S^{-\tau}$, for the distribution of avalanche sizes, $S$; 
{\bf (x)} $P(T)\sim T^{-\tau_t}$, for durations, $T$; 
and (xi) $\langle S \rangle \sim T^\gamma$ linking durations with averaged sizes, $\langle S \rangle$.

Note that these last exponents (spreading and avalanche ones) are not independent of each other, but related through scaling relations; e.g.~\citep{munoz1999}:
\begin{eqnarray}
\tau &=& \frac{1 + \eta + 2\delta}{1 + \eta + \delta} \; , \label{eq:pred_tau}\\
\tau_t &=& 1 + \delta \; , \label{eq:pred_tau_t} \\
\gamma &=&  \frac{\tau_t - 1}{\tau - 1} = 1 + \delta + \eta \; , \label{crackling}
\end{eqnarray}
where the last one describes the ``crackling noise'' scaling relation~\citep{Crackling_Sethna}. Other scaling relations can be found in ~\citep{munoz1999,Odor,Henkel}, in particular, 
\begin{equation}
\theta = \beta/\nu_\parallel \; ,\label{eq:scaling}
\end{equation}
relates static and dynamic exponents.

Associated with the crackling noise exponent, for standard processes with absorbing states (e.g., DP and TDP), the averaged shape of avalanches with different durations and sizes (or ``mean temporal profile of avalanches") collapses onto a universal curve that typically has a symmetric parabolic form (see Sec.~\ref{sec:MTP})~\citep{Friedman2012,Serena-BP}).

It is noteworthy that there is a set of exponents that can be argued to remain unchanged across transition types (a fact that is also confirmed numerically). 
Due to the diffusive nature of the system in all continuous transitions, correlations ($\xi$) should diverge at the critical point with mean-field exponents ($\nu$) as follows:
$\xi_\perp\propto \Delta^{\nu_\perp}$ with $\nu_\perp=1/2$, for the correlation length;
and $\xi_\parallel \propto \Delta^{\nu_\parallel}$ with $\nu_\parallel = 1$, for the time correlation. 
From this, given that~\citep{munoz1999} $z = 2\nu_\perp/\nu_\parallel = 1$, $z = 1$ for all continuous transitions here.
Similarly, the survival probability exponent (whose scaling behavior was determined in~\citep{Survival1997}) always takes a value $\delta = 1$ for all the continuous transitions studied here, implying that $\tau_t = 2$ (see Eq.~\eqref{eq:pred_tau_t}) is conserved across transitions.
Finally, the exponent $\eta$ is expected to vanish for all mean-field transitions (for which there is no ``anomalous dimension"~\citep{GGMA}).
However, remarkably, here we report on a possible exception to this general rule ($\eta=2$) for one of the ``anomalous" transitions.

\begin{figure*}[!t]
\includegraphics[width = \textwidth]{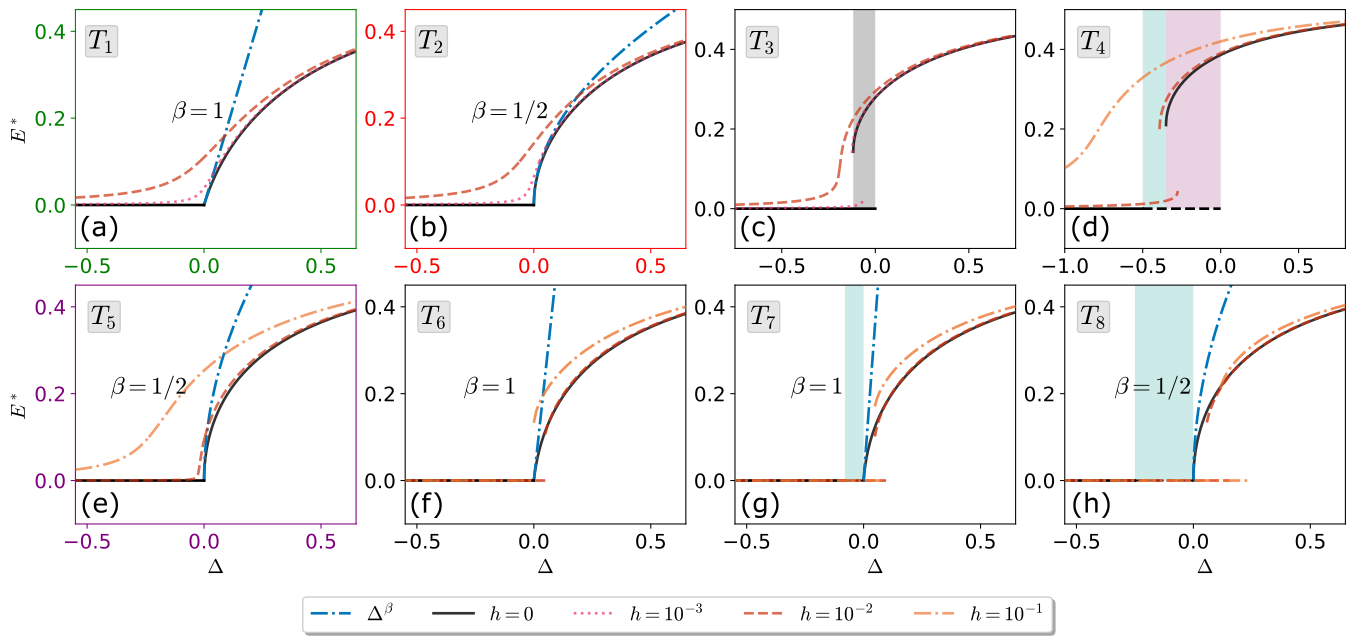}
\caption{ 
Order parameter as a function of the control parameter, $\Delta = w_{EE} - w_{EE}^*$, across the eight possible types of transitions as represented in Fig~\ref{fig:phase-diagram}.
In all plots revealing continuous transitions, the dot-dashed blue curves correspond to asymptotic behavior $E^*(\Delta;h=0)\sim\Delta^{\beta}$, with the corresponding values of the $\beta$ exponent.
The shaded grey areas correspond to bistability between active and standard quiescent phases and the pink shaded area between active and excitable quiescent phases.
The green shaded areas represent the excitable quiescent phase (same colors as Fig.~\ref{fig:phase-diagram}).
(a) At $T_1$ ($w_{IE} = 3$), the system exhibits a second-order transition from the standard quiescent to the active phase, consistently with the directed-percolation (DP) universality class ($E^* \sim \Delta^1$ for $\Delta \geq 0$).
(b) $T_2$ ($w_{IE} = 3$) is also a continuous phase transition occurring through a tricritical point and is consistent with the tricritical directed percolation (TDP) universality class ($E^*\sim\Delta^{1/2}$).
(c, d) In contrast, $T_3$ and $T_4$ are first-order or discontinuous phase transitions with coexistence between an active phase and one of two possible kinds of quiescence ($w_{IE} = 3$): first, a standard quiescent state (grey shaded area) and second an excitable quiescent state (pink shaded areas).
(e) Case B ($w_{IE} = 1$) allows for a special tricritical transition ($T_5$) occurring through a Hopf-tricritical point ($E^*\sim\Delta^{1/2}$).
(f, g, h) In case C ($w_{IE} = 0.8$), both $T_6$ and $T_7$ are continuous phase transitions when $h=0$, with $E^*\sim\Delta^1$ and $T_8$, $E^*\sim\Delta^{1/2}$, respectively.
However, once a non-vanishing external field $h \neq 0$ is introduced (dash-dotted and dashed lines), there is bistability driven by the external field (for more details see grey shaded areas in Figs.~\ref{fig:powerlaws_external_field}f-h).
Parameter values are set as in Fig~\ref{fig:phase-diagram}.
}
\label{fig:beta_exponents}
\end{figure*}

\begin{table}[t]
\caption{Summary of mean-field exponents ~\citep{munoz1999, Lubeck06}.}
\centering
\begin{tabular}{l c c c c c}
\\[-2ex]\hline 
     \hline \\[-2ex]
 & DP & & TDP &  & H+TDP \\
\hline
Codim. & 1 && 2 && 3\\
\hline
\hline
$\boldsymbol\beta$ & 1 && 1/2 && 1/2 \\ 
$\boldsymbol\delta_h$ & 2 && 3 && 2 \\
$\boldsymbol\theta$ & 1 && 1/2 && 1 \\ 
$\boldsymbol\delta$ & 1 && 1 && 1 \\ 
$\boldsymbol\eta$ & 0 && 0 && 2 \\
$\nu_\parallel$ & 1 && 1 && 1\\
$\boldsymbol\tau$ & 3/2 && 3/2 && 5/4 \\ 
$\boldsymbol\tau_t$ & 2 && 2 && 2 \\
$\boldsymbol\gamma$ & 2 && 2 && 4 \\
\\[-2ex]\hline 
     \hline \\[-2ex]
\end{tabular}
\label{tab:exponents}
\end{table}

\begin{figure*}[!t]
\centering
\includegraphics[width = \textwidth]{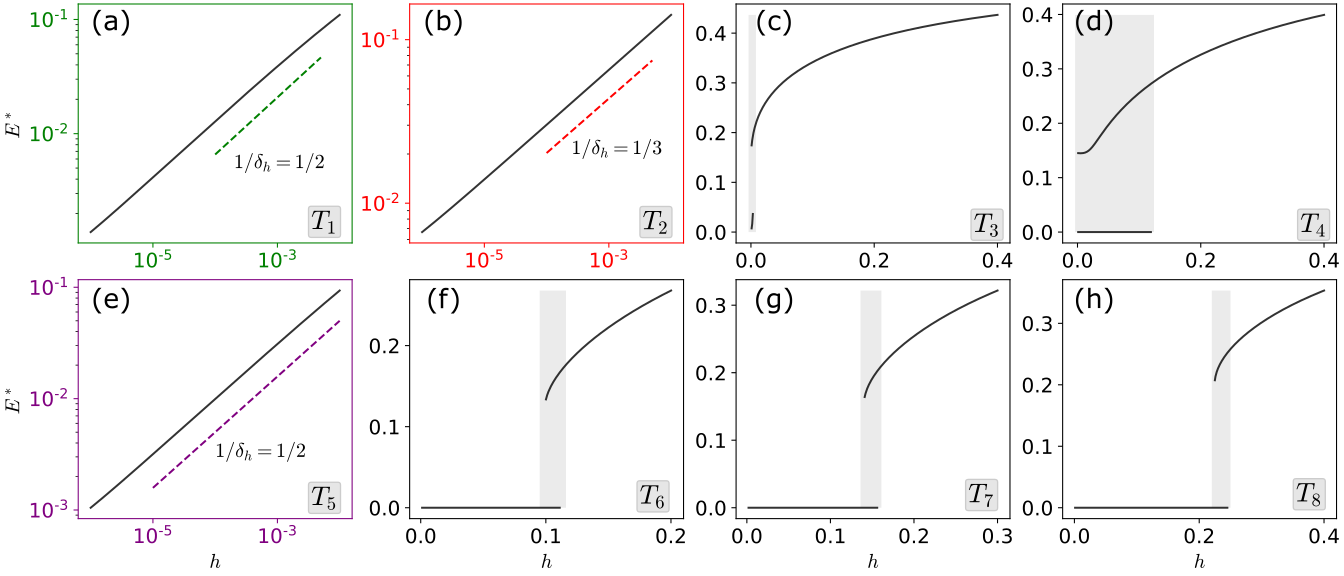}
\caption{Order parameter as a function of the external field right at the transition ($\Delta = 0$). 
Observe that three out of the eight types of transitions described here exhibit power law scaling with the external field, i.e., $E^*(h;\Delta = 0)\sim h^{1 / \delta_h}$. 
(a) For $T_1$, $\delta_{h} = 2$, consistently with the DP universality class.
(b) For $T_2$, $\delta_{h} = 3$, consistently with TDP.
(c, d) For $T_3$ and $T_4$, the order parameter's response to an external field shows bistability (shaded area).
(e) Transition $T_5$ (H+TDP), differently from the usual tricritical transition (TDP), scales with $\delta_h = 2$.
(f, g, h) Remarkably, for $T_6$, $T_7$, and $T_8$, contrary to the behavior with $h=0$, the order parameter becomes bistable (shaded area) as the external field increases. 
Parameter values as in Fig.~\ref{fig:phase-diagram}.
}
\label{fig:powerlaws_external_field}
\end{figure*}

\subsection*{\texorpdfstring{$T_1$: Directed Percolation\\ (Transcritical bifurcation)}{TEXT}}

 $T_1$  corresponds to a transcritical bifurcation, describing a continuous transition between the standard quiescent and active phases.
As discussed in Section~\ref{sec:intro}, guided by universality principles, one expects it to lie in the usual (mean-field) directed percolation universality class (DP)~\cite{Janssen,Grassberger,Marro,Henkel,GGMA}.
Indeed, this is the case, as explicitly shown in what follows.

Transcritical bifurcations occur when the quiescent steady state loses its local stability, i.e., Eq.~\eqref{eq:transc_wEE}. 
Expanding Eq.~\eqref{eq:WCEI-Eeq} and Eq.~\eqref{eq:WCEI-Ieq} in power series of $E^*$ and $I^*$, one finds:
\begin{equation}
E^*(\Delta; h = 0) = \frac{(\alpha + w_{II})^3}{(\alpha + w_{II})^3 - w_{EI}w_{IE}^2}\Delta + {\cal O}(\Delta^2)\; , \label{eq:Transc_E*_delta}
\end{equation}
from which $\beta = 1$ follows.
The introduction of an external field $h$ smooths out the transition (as illustrated with dashed lines in Fig.~\ref{fig:beta_exponents}a). 
Hence, expanding the fixed point in powers of $h$, at $\Delta=0$, yields:
{\medmuskip=0mu
\thinmuskip=0mu
\thickmuskip=0mu
\begin{equation}
E^*(h;\Delta=0) = \sqrt{\frac{(\alpha + w_{II})^2(w_{EI} - \alpha)h}{\alpha\left[w_{IE}w_{EI}^2 - (w_{II} + \alpha)^3\right]}} + {\cal O}(h)\; ,
\label{eq:Transc_E*_h}
\end{equation}
}
\!\!so that $\delta_{h} = 2$ (see Fig.~\ref{fig:powerlaws_external_field}a).

Similarly, one can derive the solution for $I(t)$ and, by expanding it in a power series, obtain \mbox{$I(t)\approx[w_{IE}/(w_{II}+\alpha)]E(t)$}.
It is, thus, convenient to define two new variables: $\Sigma$ and $\Lambda$, as the weighted linear combinations:
\begin{eqnarray}
2\Sigma &=& w_{IE}\;E + (w_{II} + \alpha)I \; ,\\
2\Lambda &=& w_{IE}\;E - (w_{II} + \alpha)I\; ,
\end{eqnarray}
in terms of which the mean-field dynamics (Eq.\eqref{eq:WCEI-Eeq} and Eq.~\eqref{eq:WCEI-Ieq}) is rewritten in a simpler form:
{\medmuskip=0mu
\thinmuskip=0mu
\thickmuskip=0mu
\begin{eqnarray}
2\dot\Sigma &\approx& \Delta\Sigma + (2w_{EE} + 2w_{II} + \Delta)\Lambda + {\cal O}\label{eq:dotSum}\\
2\dot\Lambda &\approx& \Delta\Sigma - \left[2(w_{EE} + w_{II} - 2\alpha) + \Delta\right]\Lambda + {\cal O}\; , \label{eq:dotLambda}
\end{eqnarray}
}
where $ {\cal O}={\cal O}(\Sigma^2, \Lambda^2, \Sigma\Lambda, ...)$ stands for higher-order terms.
Observe that, right at the transition ($\Delta=0$), the stability matrix around the origin is 
\begin{eqnarray}
A = \left(
\begin{array}{c c}
0 & w^T_{EE} + w_{II}\\
0 & w^T_{EE} - w_{II} - 2\alpha\\
\end{array}
\right)\; ,
\end{eqnarray}
which has only one vanishing eigenvalue, while the second one is strictly negative at criticality for $T_1$ transitions.
This means that $\Lambda$ decays exponentially fast, therefore, it is an \emph{irrelevant} field for scaling.
Only one ``slow mode'' or ``relevant field'' exists, $\Sigma$, and --- as theoretically predicted in Grinstein {\emph{et al.}}~\citep{Grinstein1989} for these conditions --- the scaling behavior should coincide with standard DP.

In particular, at the transition point ---where the linear term of Eq.~\eqref{eq:dotSum} vanishes--- the quadratic term dominates and therefore $\Sigma(t) \propto t^{-1}$, so that  $\theta = 1$ (as numerically confirmed in Fig.~\ref{fig:powerlaws_time_decay}a).
Considering the previous three independent exponent values, one can already conclude that the $T_1$ transition actually belongs  in the DP universality class (see Table~\ref{tab:exponents}). 
Nevertheless, for the sake of completeness, the survival probability and the total number of particles, at $\Delta = 0$, are confirmed to scale with spreading-exponent values  $\delta = 1$ (Fig.~\ref{fig:survival_prob}a) and $\eta = 0$ (Fig.~\ref{fig:eta_exponents}a), respectively, as expected for the DP class.
We have also confirmed the consistency with DP by numerically analyzing the statistics of avalanches at the transition, revealing exponent values compatible with the DP predictions $\tau=3/2$, $\tau_t=2$, and $\gamma=2$ (see the distributions of sizes $S$, durations $T$, and average sizes as a function of durations in Figs.~\ref{fig:avalanches}a, \ref{fig:avalanches}d, and~\ref{fig:avalanches}g, respectively).
Moreover, the averaged avalanche shape is approximately an inverted parabola throughout the $T_1$ line (Figs.~\ref{fig:Skw_tempprofile}a and~\ref{fig:Skw_tempprofile}b), collapsing for different durations with $\gamma = 2$, even if with some asymmetry (see Section~\ref{sec:MTP} for a more in-depth discussion on avalanche shapes).

Thus, in summary, at the line of transcritical bifurcations ($T_1$) separating a standard quiescent from the active phase, the Wilson-Cowan stochastic model exhibits a genuine critical point in the DP class, a result that is consistent with recent analyses of de Candia~{\it et al.}~\citep{deCandia21} for their specific choice of parameter values.

\subsection*{\texorpdfstring{$T_2$: Tricritical Directed Percolation \\ (Saddle-node-transcritical bifurcation)}{TEXT}}
\label{sub:SNT}

The tricritical point in case A (see Fig.~\ref{fig:phase-diagram}a) corresponds to a saddle-node transcritical (SNT) bifurcation --- i.e., where the lines of transcritical and saddle-node bifurcations intersect without further degeneracies~\citep{SNT2019,SNT2020}.
Thus, in order to tune to this transition point one needs to set two parameters in the phase diagram ($w_{EE}$, $w_{EI}$), as explicitly calculated in Appendix~\ref{ap:mat_TDP}.
An analysis in terms of the fields $\Sigma$ and $\Lambda$ (similar to the previous case) shows that there is only one vanishing eigenvalue at the transition, and, thus, the second field is irrelevant for scaling.
Therefore, $T_2$ is expected to be described by the mean-field tricritical directed percolation universality class (TDP)~\citep{Lubeck06}.
Indeed, considering the leading-order term in a power expansion in both $\Delta$ and $h$, one has:
{\medmuskip=0mu
\thinmuskip=0mu
\thickmuskip=0mu
\begin{eqnarray}
E^*(\Delta, h=0) &\approx& \sqrt\frac{\Delta}{w_{IE}} + {\cal O}(\Delta),
\label{eq:rho_exd_TDP}\\
E^*(h,\Delta=0) &\approx&
\left[\frac{3\left[w_{IE}^2 - (\alpha + w_{II})^2\right]h}{w_{IE}^2\left[(\alpha^2 - 3)w_{IE} - (\alpha^2 + 3)\alpha\right]}\right]^{\frac{1}{3}} + {\cal O}\left(h^\frac{1}{2}\right) \nonumber \\ \; 
\label{eq:rho_exh_TDP}
\end{eqnarray}
}
\!\!from where $\beta = 1/2$ (Fig.~\ref{fig:beta_exponents}b) and $\delta_{h} = 3$ (Fig.~\ref{fig:powerlaws_external_field}b), as expected for the TDP universality class.

\begin{figure*}[!t]
\centering
\includegraphics[width = \textwidth]{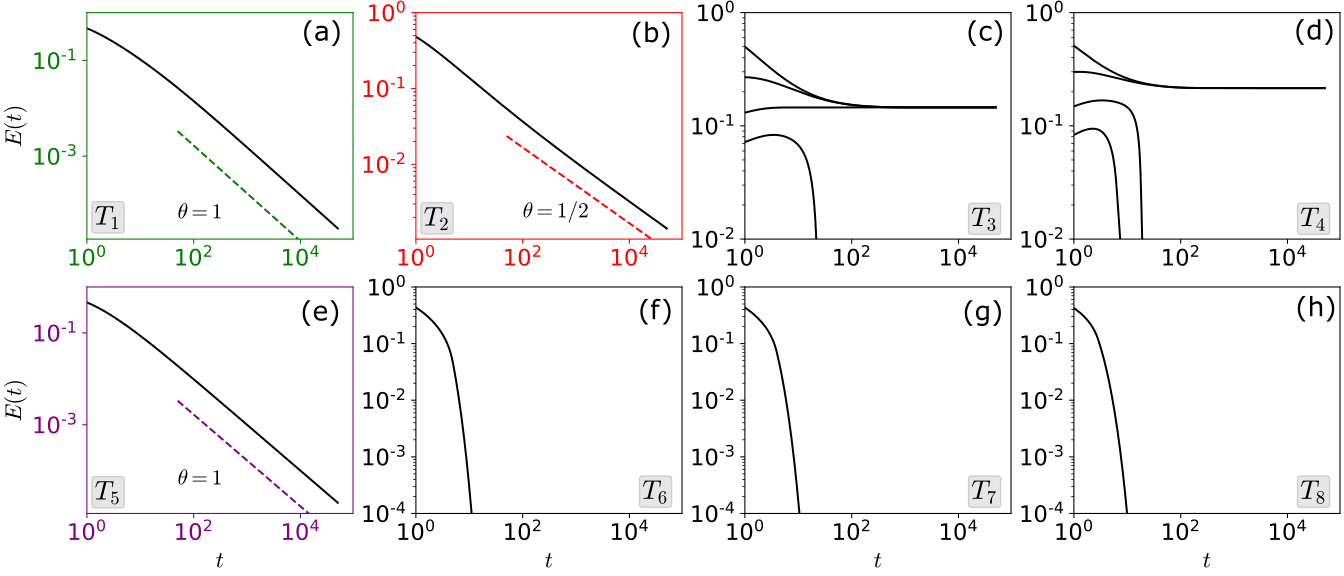}
\caption{Order parameter time series for the eight types of transitions. 
Observe that only three of them exhibit dynamical scaling $E(t) \propto t^{-\theta}$.
(a) $T_1$ exhibits an asymptotic power law decay with the expected DP value $\theta = 1$.
(b) $T_2$ shows a slower asymptotic time decay in the TDP class, $\theta = 1/2$.
(c) For $T_3$, the saddle-node bifurcation gives rise to bistability between an active and a quiescent phase.
(d) $T_4$ behaves very similarly to $T_3$ but frustrated oscillations drive the system more easily to regions II and III, so the bistability is between active and excitable quiescent phases.
(e) $T_5$ is a genuine second-order phase transition with $\theta = 1$.
(f,g,h) $T_6$, $T_7$, and $T_8$ are not genuine continuous transitions and show no signatures of dynamic scaling, but rather an exponential decay to quiescence.
Parameter sets as in Fig.~\ref{fig:phase-diagram}.
}
\label{fig:powerlaws_time_decay}
\end{figure*}

At the transition, the lowest order correction of Eq.~\eqref{eq:dotSum} in $\Sigma$ is ${\cal O}(\Sigma^3)$, so that asymptotically $\Sigma \propto t^{-1/2}$ and, hence, $\theta = 1/2$, as numerically confirmed in Fig.~\ref{fig:powerlaws_time_decay}b. 
Once again, considering the linear relationship between $E(t)$ and $I(t)$, both densities share this scaling.

Finally, the exponent for the survival probability remains $\delta = 1$ (see Fig.~\ref{fig:survival_prob}b), $\eta=0$  (see Fig.~\ref{fig:eta_exponents}), $\tau = 3/2$ (Fig.~\ref{fig:avalanches}b), $\tau_t = 2$ (Fig.~\ref{fig:avalanches}e), and $\gamma=2$ (Fig.~\ref{fig:avalanches}h), all of which are consistent with the expected values in the TDP class (see Table~\ref{tab:exponents}).

\subsection*{\texorpdfstring{$T_3$: Standard discontinuous transition \\ (Saddle-node bifurcation)}{TEXT}}

The line of saddle-node bifurcations (see Fig.~\ref{fig:phase-diagram}a), Eq.~\eqref{eq:w_EE_function}, defines the third type of transition, $T_3$, to go from a standard quiescent state to the active phase.
This type of transition is characterized by a discontinuous jump in the order parameter and includes an intermediate regime of bistability, where both the active and the standard quiescent state are stable (see Figs.~\ref{fig:phase-diagram}a and~\ref{fig:beta_exponents}c).
The regime of coexistence lasts until, at a second bifurcation, the quiescent phase loses its local stability.
Given that the transition is discontinuous, the exponents $\beta$ and $\delta_h$ are not properly defined (Fig~\ref{fig:powerlaws_external_field}c).
Similarly, neither the activity nor the survival probability decay to $0$ for initial conditions in the basin of attraction of the active phase (see Figs.~\ref{fig:powerlaws_time_decay}c and~\ref{fig:survival_prob}c), so that the exponents $\theta$ and $\delta$ are not well-defined either.

Thus, in summary, the $T_3$ transition is just a standard first-order or discontinuous transition into a quiescent state~\citep{Lubeck06,Odor}.

\subsection*{\texorpdfstring{$T_4$:  Discontinuous transition from an excitable quiescent state \\ (Saddle-node bifurcation)}{TEXT}}

\begin{figure*}[!t]
\centering
\includegraphics[width = \textwidth]{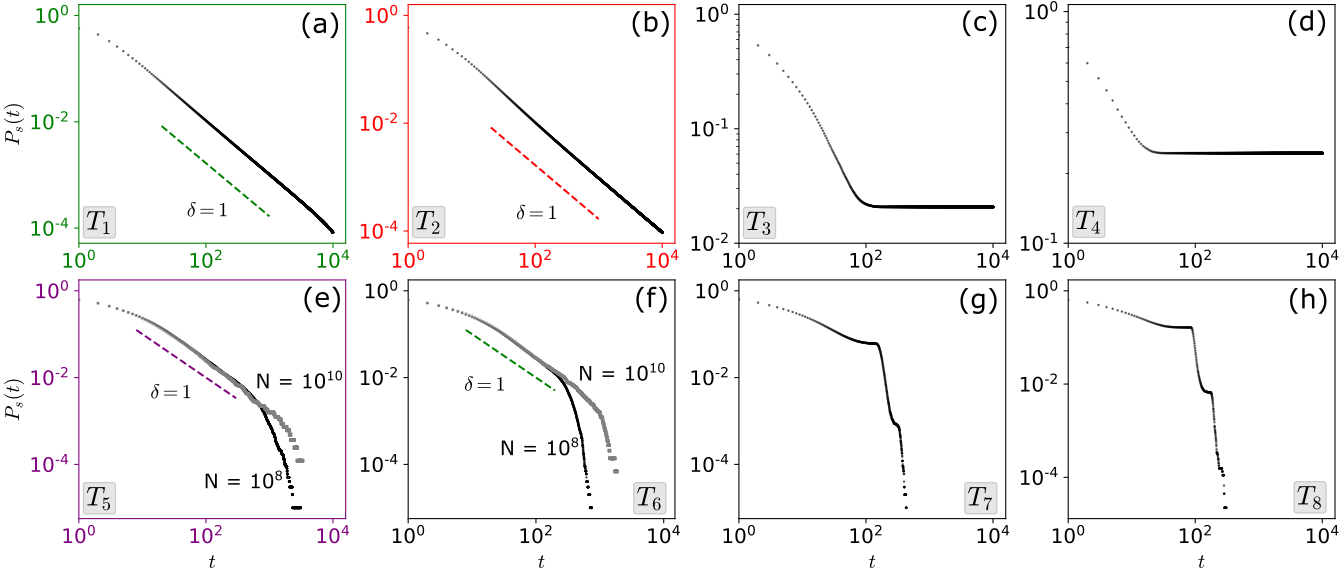}
\caption{Survival probability as a function of time.
The survival probability at second-order phase transitions scales as $P_s(t)\propto t^{-\delta}$. 
Black dots stand for the numerical simulations (for the same parameters as Fig.~\ref{fig:phase-diagram} and $N=10^8$) and dashed lines show the corresponding exponent value.
(a) $T_1$ belongs to the DP universality class, i.e., $\delta = 1$. 
(b) $T_2$ belongs to TDP so that $\delta = 1$ (same as DP).
(c, d) For the first-order phase transitions, the system's survival probability converges to a non-vanishing value as $t\rightarrow \infty$ due to the possibility of being attracted to the active phase.
(e) For $T_5$, $\delta = 1$ as in the previous continuous transitions, but with stronger finite-size effects.
(f) For $T_6$, the system shows a behavior similar to $T_5$: a decay with $\delta = 1$ and strong finite-size effects. 
(g, h) The survival probability shows a peculiar behavior of several sharp decays with some small plateaus, which stem from the excitability of the quiescent phase.}
\label{fig:survival_prob}
\end{figure*}

A scenario very similar to $T_3$ occurs at $T_4$, which appears in all three possible phase diagrams (A, B, and C; see Fig~\ref{fig:phase-diagram}).
Transition $T_4$ is also discontinuous with phase coexistence, but it differs from $T_3$ in the fact that --- as illustrated in Fig.~\ref{fig:beta_exponents}d --- the quiescent phase that coexists with the active one in the regime of bistability is of the excitable type, rather than the standard one.
For the same reasons as in $T_3$, none of the critical exponents is well defined (see Figs.~\ref{fig:powerlaws_external_field}d, \ref{fig:powerlaws_time_decay}d, and~\ref{fig:survival_prob}d).

Thus, in summary, $T_4$ is a discontinuous transition with bistability, but with the peculiarity of having an excitable quiescent state coexisting with the active one.

\subsection*{\texorpdfstring{$T_5$: Hopf Tricritical Directed Percolation\\  (Hopf saddle-node-transcritical bifurcation)}{TEXT}}\label{sec:SNTAH}

As illustrated in Fig.~\ref{fig:phase-diagram}b, Case B exhibits a \emph{codimension 3} bifurcation point at which the tricritical point (codimension 2) and the line of Hopf bifurcations meet.
This transition occurs only in case B, for the particular choice of parameters for which the vertical line of Hopf bifurcations ends up exactly at the tricritical point, $w^*_{IE} = \alpha + w_{II}$, as derived from Eq.~\eqref{eq:AHT_wei} and Eq~\eqref{eq:SNT_wei_A} in Appendix B. 
Using this constraint, one can easily find that the location of the $T_5$ point is specified by the following set of conditions
(see Appendix B):   
$w_{EE} = 2\alpha + w_{II}$ and $
w_{EI} = w^*_{IE}$.

Let us first write the stationary solutions of the dynamical equations, Eq.~\eqref{eq:WCEI-Eeq} and Eq.~\eqref{eq:WCEI-Ieq}, up to leading order in $\Delta$ at vanishing $h$ and, also,  up to leading order in $h$ at vanishing $\Delta$, i.e., 
\begin{eqnarray}
E^*(\Delta; h = 0) &\approx& \sqrt\frac{\Delta}{\alpha + w_{II}} + {\cal O}(\Delta)\;, \label{eq:E_FP_delta_TDPA}\\
E^*(h; \Delta = 0) &\approx& \sqrt{\frac{h}{3(\alpha + w_{II})^2}} + {\cal O}(h)\;.  
\end{eqnarray} 
These imply $\beta = 1/2$ (as illustrated in Fig.~\ref{fig:beta_exponents}e) and $\delta_h=2$ (see Fig.~\ref{fig:powerlaws_external_field}e).
Note that $\beta$ coincides with its counterpart for TDP (as expected for a tricritical-like point) but, curiously enough, $\delta_h$ does not; 
it instead coincides with its value in the DP class.
Therefore, the static exponents at $T_5$ do not fully comply with either of the well-known universality classes.

To make further progress, it is convenient to write the equations for $\Sigma(t)$ and $\Lambda(t)$ for case B, as defined in Eq.~\eqref{eq:dotSum} and Eq.~\eqref{eq:dotLambda}:
    \begin{eqnarray}
        2\dot\Sigma (t) &=& \Delta(\Sigma + \Lambda) + 4(w_{II} + \alpha)\Lambda  - \frac{2\alpha}{w_{II}+\alpha}\Sigma^2
        - 4\Sigma\Lambda
        \nonumber\\ 
        &&  -\frac{2\alpha}{w_{II}+\alpha}\Lambda^2
    + {\cal O} \label{eq:niceq1} \\
        2\dot\Lambda(t) &=& \Delta(\Sigma + \Lambda)  - 4\Lambda^2 - \frac{4\alpha}{w_{II}+\alpha}\Sigma\Lambda + {\cal O}  
       \label{eq:niceq2}
    \end{eqnarray}
    where ${\cal O}\equiv{\cal O}(\Sigma^3, \Sigma^2\Lambda, \Sigma\Lambda^2, \Lambda^3, \Delta\Sigma^2...)$ stands for higher-order terms and time dependences have been omitted for simplicity.
Moreover, right at the transition point ($\Delta = 0$),  the dynamics simplifies to
{\medmuskip=0mu
\thinmuskip=0mu
\thickmuskip=0mu
\begin{eqnarray}
\dot\Sigma (t) &=& 2(w_{II} + \alpha)\Lambda - \frac{\alpha}{(w_{II} + \alpha)}\Sigma^2 + {\cal O}
\label{eq:sigma_TDPA}\; \\
\dot\Lambda (t) & = & 
  - 2 \Lambda^2 -\frac{2\alpha}{w_{II} + \alpha}\Lambda\Sigma + ~{\cal O}
\; . \label{eq:lambda_TDPA}
\end{eqnarray}}
where the consistency of the truncation of higher-order terms will be justified a posteriori.

In particular, observe that  
the stability matrix around the origin becomes
\begin{eqnarray}
A \propto \left(
\begin{array}{c c}
0 & w_{II} + \alpha\\
0 & 0\\
\end{array}
\right)\label{eq:mat_BT}
\end{eqnarray}
so that the null eigenvalue is degenerate and, thus, an anomalous type of scaling is to be expected. 
Indeed, the previous matrix is characteristic of a Bogdanov-Takens bifurcation~\citep{Izhi}, which has been already discussed in the context of Wilson-Cowan models~\cite{WC-review,Corral22} and, more in general, in the analysis of non-normal or \emph{non-reciprocal phase transitions}~\cite{Non-reciprocal}.

It is important to observe that the only linear term in the first equation, Eq.\eqref{eq:sigma_TDPA}, $2 (w_{II} + \alpha)\Lambda(t)$, has a positive coefficient. This implies that at criticality $\Lambda(t)$ needs to decay to zero faster than $\Sigma(t)$ as otherwise the overall right-hand-side would be positive asymptotically in time (which cannot possibly happen at criticality). 
Therefore, given that $\Sigma(t)$ needs to decay slower than $\Lambda$, the slowest-decaying non-linear term in Eq.\eqref{eq:sigma_TDPA} is the one proportional to $-\Sigma^2$.
Knowing  that asymptotically, $\dot\Sigma \propto - b \Sigma^2$, with $b=\alpha/(w_{II} + \alpha)$ one readily finds that $\Sigma \sim t^{-1}/b$ and, therefore, $\theta=1$ (Fig.~\ref{fig:powerlaws_time_decay}e). 
Finally, plugging this  result into the second equation, Eq.\eqref{eq:lambda_TDPA}, comparing constants and exponents, one readily finds that 
$\Lambda \sim t^{-2}$. 
Observe that, indeed, as anticipated, $\Lambda(t)$ decays faster than $\Sigma(t)$: $\Lambda \sim \Sigma^2$, which justifies the truncation of higher-order terms in  the previous equations.

Using these observations one concludes that, right at the transition point $\Delta = 0$, the dynamical scaling is consistent with DP because, since $\Lambda$ decays faster, it does not influence the decay of $\Sigma$ (dominated by a quadratic term). 
This result is surprising  as we are dealing with a tricritical point so, a priori, one would expect TDP-like scaling.

The situation is different away from the critical point ($\Delta > 0$). 
In this case, it is convenient to focus on the equation for $\dot \Lambda$, Eq.~\eqref{eq:niceq2}. 
At stationarity, the linear positive term (proportional to $\Delta$) needs to cancel with the leading non-linear one.
A priori, the linear positive term is either the one proportional to $\Delta \Sigma$ or the one proportional to $\Delta \Lambda$, depending on the scaling dimensions of $\Sigma$ and $\Lambda$. 
Note that both yield that $\Lambda$ scales as $\Lambda \propto \Delta$.
Now, focusing on the first equation (Eq.~\eqref{eq:niceq1}), the leading positive term is $4(w_{II}+ \alpha) \Lambda$ (which scales as $\Delta$, while $\Delta (\Sigma + \Lambda)$ is a higher-order contribution). 
This leading term needs to be comparable with the leading negative term, which is the one proportional to $-\Sigma^2$ (note that the  other possibility, $-4 \Sigma \Lambda$, leads to a fixed value of $\Sigma$ that does not change/scale with $\Delta$ and, therefore, it is not a solution).
The resulting scaling renders $\Lambda \sim \Sigma^2$, which is consistent with the temporal scaling.
And, then, one derives $\Sigma \sim \Lambda^{1/2} \sim \Delta^{1/2}$, i.e., $\beta=1/2$ (while the field $\Lambda$ scales with an exponent $\beta_{\Lambda}=1$).

Therefore, since 
(i) the order-parameter exponent is that of the TDP class, $\beta=1/2$, 
(ii) the time-decay exponent $\theta=1$ differs from its TDP value, and 
(iii) $\nu_\parallel = 1$ (as it is the case for all mean-field transitions), then it follows that
\begin{equation}
\theta \neq \beta/\nu_\parallel \; ,\label{eq:sclbreak}
\end{equation}
which violates one of the basic scaling relations in systems with quiescent states, i.e., Eq.(\ref{eq:scaling}).

Let us remark that a similar violation of scaling was found by Noh and Park~\citep{NohPark05} in a model with quiescent states and two relevant fields: an ``excitatory" and a ``repressing" one. 
In both cases ---here and \cite{NohPark05}--- the breakdown of scaling stems from the non-trivial interplay between these two fields with opposing effects. 

Similarly to the other transitions, the survival probability decays with an exponent $\delta = 1$, albeit with a higher sensitivity to system size (see Fig.~\ref{fig:survival_prob}e).
Also, consistently with the scaling relation $\tau_t = \delta + 1$~\citep{munoz1999}, the avalanche distribution of durations scales with the same exponent as in DP and TDP, $\tau_t \approx 2$ (Fig.~\ref{fig:avalanches}f).

\begin{figure}[t!]
\centering
{\includegraphics[width = \columnwidth]{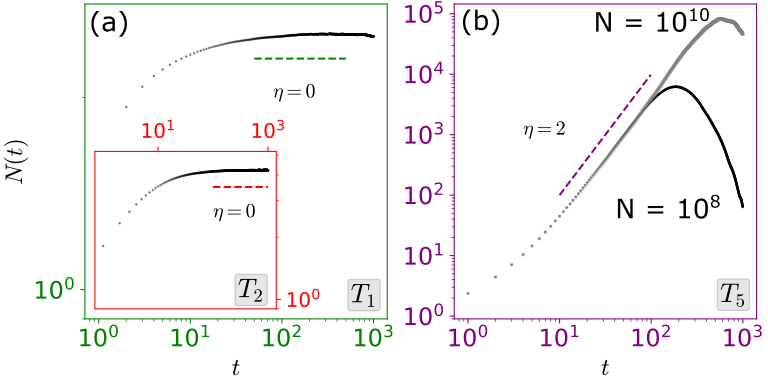}}
\caption{Mean number of particles $N(t)$ in spreading experiments in bonafide continuous phase transitions (i.e., $T_1$, $T_2$, and $T_5$), at which one expects $N(t)\sim t^\eta$.
Simulations with the same parameters as Fig.~\ref{fig:phase-diagram} with $N=10^8$ [panel (a)] and $N = 10^8$ and $N= 10^{10}$ [panel (b)].
(a) For $T_1$ and $T_2$, we obtain results compatible with $\eta = 0$, as expected for DP and TDP as well as, in general, for mean-field theories.
(b) On the other hand, for $T_5$ we obtain the unusual result $\eta = 2$, with strong finite-size effects.
}
\label{fig:eta_exponents}
\end{figure}

\begin{figure*}[!t]
\centering
\includegraphics[width = 0.82\textwidth]{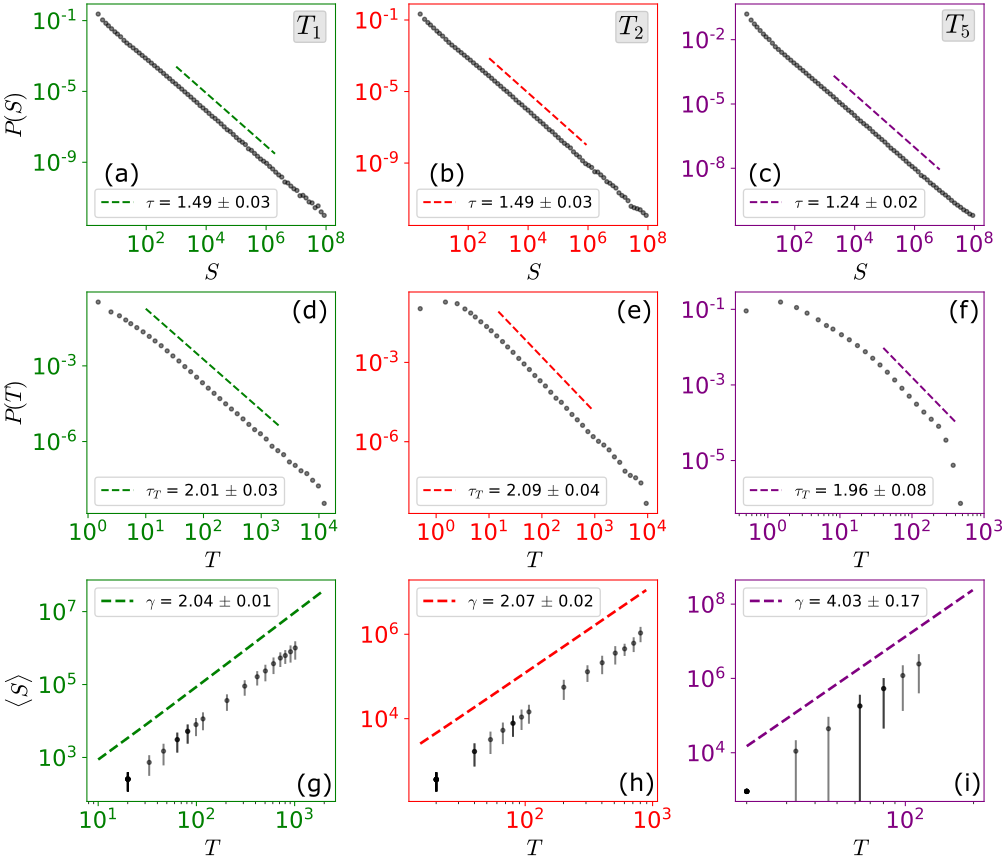}
\caption{ {Avalanche analysis.} Simulations using Gillespie's algorithm for the same parameters as in Fig.~\ref{fig:phase-diagram} and network size $N=10^8$.
In this case, we report results only for true (or bonafide) continuous phase transitions, for which scale-free avalanches emerge, i.e., $T_1$, $T_2$, and $T_5$.
For the $T_1$ transition, one obtains results as expected for DP: (a) $\tau\approx 3/2$, (d) $\tau_t\approx 2$, and (g) $\gamma \approx 2$.
For $T_2$, the system behaves consistently with TDP: 
(b) $\tau\approx 3/2$, (e) $\tau_t\approx 2$, and (h) $\gamma \approx 2$, i.e., TDP and DP share the same avalanche exponents.
Finally, for the $T_5$ transition, (c) $\tau \approx 5/4$, (f) $\tau_t \approx 2$, and (i) $\gamma \approx 4$.
}
\label{fig:avalanches}
\end{figure*}

In contrast with the rest of the second-order phase transitions (see Fig.~\ref{fig:eta_exponents}a), the growth of the total activity in spreading experiments right at criticality is $N(t) \sim t^{2}$, yielding $\eta = 2$ (Fig.~\ref{fig:eta_exponents}b).
This observation is rather surprising for a mean-field model as most mean-field universality classes are characterized by $\eta=0$ (i.e., absence of an ``anomalous dimension"~\citep{Binney,GGMA}).  

In order to shed some light on this result, let us observe that the linearized dynamics at criticality --- controlled by the normal form of a Bogdanov-Takens bifurcation, Eq.~\eqref{eq:mat_BT} --- is such that a small initial perturbation can be largely amplified before decaying back to quiescence, i.e., around the fixed point the system is excitable.
In particular, if the perturbation consists of a single seed (as in spreading experiments), the number of active sites in surviving runs grows in a deterministic way until a maximum size is reached, and, then, the asymptotic decay toward the quiescent state (controlled by the exponent $\theta$) begins.
This initial deterministic growth --- which does not occur in the DP nor TDP classes --- is expected to be responsible for the anomalous value of $\eta$.

More specifically, observe that the density $\Sigma$ at first grows linearly --- as $\dot{\Sigma} \propto\Lambda$ and $\Lambda$ can be approximately taken as a constant because its negative eigenvalue vanishes [Eq.~\eqref{eq:sigma_TDPA} and Eq.~\eqref{eq:lambda_TDPA}]. 
Furthermore, the total number of active sites $N(t)$ is equal to the density times an additional ``volume factor", which, owing to the deterministic expansion, grows linearly in time.
Therefore, one concludes that $N(t) \sim \Sigma(t)  t \sim t^2$, which yields $\eta=2$.

Given this anomalous value and using the general scaling relations described before, one can infer other exponent values. 
In particular, Eq.~\eqref{eq:pred_tau} predicts $\tau = 5/4$ and Eq.~\eqref{crackling} leads to $\gamma = 4$, which are both unusual/anomalous exponents in mean-field theories. 
We numerically verified both of these results; scaling compatible with $\tau \approx 5/4$ can be observed in Fig.~\ref{fig:avalanches}c, and with $\gamma \approx 4$ (see Fig.~\ref{fig:avalanches}i). This latter value also gives an excellent data collapse for $P(S|T)$  (Figs.~\ref{fig:prob_cond_inset}a and~\ref{fig:prob_cond_inset}b) and is consistent with the scaling relation between size and duration cutoffs (Fig.~\ref{fig:prob_cond_inset}c)~\citep{christensen91, chessa99, DickmanCampelo03}.

In summary, the $T_5$ transition defines a thus-far unknown universality class, which we named Hopf Tricritical Directed Percolation (H+TDP).
In its mean-field variant, it has a set of exponents that do not match either DP or TDP universality classes (Table~\ref{tab:exponents}), violates at least one scaling relation, includes some anomalous exponent values, and produces highly asymmetrical avalanche shapes, as we shall show in a separate section.

A more systematic and rigorous derivation of these results, together with a field-theoretic discussion of this universality class will be presented elsewhere.

\begin{figure*}[!t]
\centering
{\includegraphics[width = \textwidth]{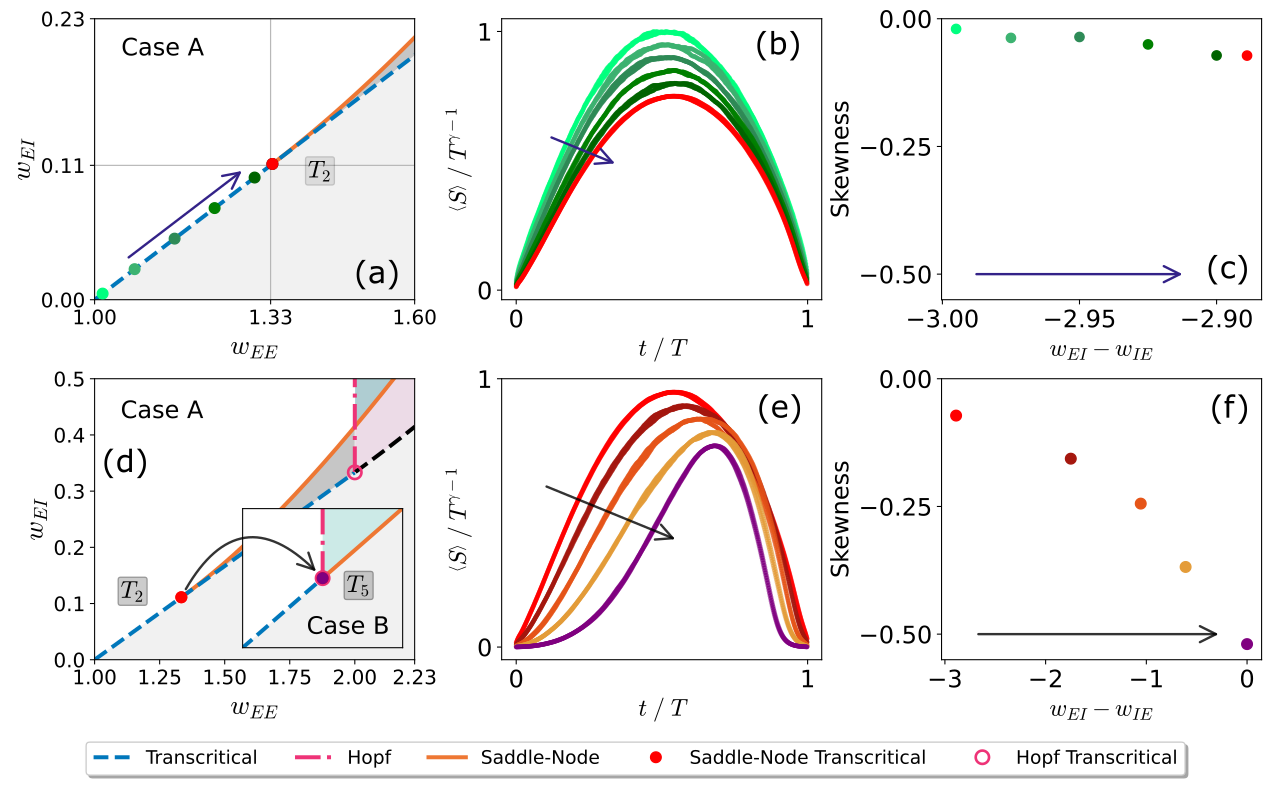}}
\caption{Skewness and mean temporal profile of avalanches.
Enlargements of Fig.~\ref{fig:phase-diagram} illustrate the excursion in parameter space (a) along a  transcritical line to a tricritical point and (d) along a tricritical line to a Hopf-tricritical point (by decreasing the parameter $w_{IE}$, the phase diagram goes from case A to case B, which is illustrated in the inset). 
(b) and (e) show the corresponding rescaled mean temporal profiles (avalanche shape collapses).
(c) The skewness of the previous curves [in panel (b)] slightly decreases as the tricritical point is approached (i.e., as the difference $w_{EI} - w_{IE}$ increases).
(e) As the system approaches the H+TDP transition, the curves of the mean temporal profile become progressively less symmetrical, as assessed by the skewness of the curves (f).
Simulations with the same parameters as Fig.~\ref{fig:phase-diagram} and $N=10^8$.}
\label{fig:Skw_tempprofile}
\end{figure*}

\subsection*{\texorpdfstring{$T_6$: Hopf-transcritical bifurcation.}{TEXT}}

This type of continuous transition (see Fig.~\ref{fig:beta_exponents}f) occurs when the line of Hopf  bifurcations collides with the transcritical line (see Fig.~\ref{fig:phase-diagram}c) and appears only in case C.
This transition is peculiar in that at the critical point --- independently of the initial conditions --- the trajectories are attracted to region II 
(and, possibly, III; see Fig.~\ref{fig:regions}b).
This occurs because the Hopf bifurcation overrides the transcritical bifurcation and the elicited frustrated oscillations drive the system toward the switching manifold and, thus, into region II.
Once in region II, the excitatory density decays exponentially fast, dragging down the system without signatures of scaling; e.g., the time-decay exponent $\theta$ is not defined for $T_6$ (see Fig.~\ref{fig:powerlaws_time_decay}f).

At the transition point (as specified by Eq.~\eqref{eq:AHT_wei} and Eq.~\eqref{eq:AHT_wee}), one can rewrite Eq.~\eqref{eq:Transc_E*_delta} and Eq.~\eqref{eq:Transc_E*_h} as:
\begin{eqnarray}
E^*(\Delta; h = 0) &\approx& \frac{\alpha+w_{II}}{(\alpha + w_{II} - w_{IE})}\Delta + {\cal O}(\Delta^2)\label{eq:delta_AHT_caseC}\\
E^*(h;\Delta = 0) &\approx& \sqrt{\frac{[(w_{II} + \alpha)^2 - \alpha w_{IE}]h}{\alpha w_{IE}[w_{IE} - w_{II} - \alpha]}} + {\cal O}(h).
\label{eq:h_AHT_caseC}
\end{eqnarray}
On the one hand, Eq.~\eqref{eq:delta_AHT_caseC} holds because, for $h=0$, there is a stable equilibrium in region I for $\Delta \geq 0$, which attracts the trajectories, preventing them from falling into region II. 
Observe that in case C, $w_{IE} < w_{II} + \alpha$ ---as the condition at the interface between case A and case C is $w_{IE} = w_{II} + \alpha$--- and, therefore, while Eq.~\eqref{eq:delta_AHT_caseC} is valid and yields $\beta=1$, Eq.~\eqref{eq:h_AHT_caseC} is misleading since the denominator is negative inside the square root so that it does not correspond to a real solution.
The reason is that the previous equations are based on the naive linearisation of the dynamical equations, assuming the non-vanishing part of the response function $\Phi$. However, this assumption is invalid in the present case. 
Thus,  for Eq.~\eqref{eq:h_AHT_caseC} and $\Delta = 0$, since trajectories fall into region II and $E$ decays to zero exponentially fast, the absorbing state remains stable even as $h$ increases from zero.
Therefore, the exponent $\delta_h$ is not well defined for $T_6$.
Nevertheless, further increasing the external field $h$ eventually leads to a saddle-node bifurcation and a discontinuity in the order parameter (Fig.~\ref{fig:powerlaws_external_field}f).

We have also verified that the system's survival probability seems to decay in time with $\delta = 1$ (Fig.~\ref{fig:survival_prob}f) as in all other transitions, even if (similarly to $T_5$) with strong finite-size effects. 
In conclusion, $T_6$ exhibits a mixture of signatures of both continuous and discontinuous transitions.

\subsection*{\texorpdfstring{$T_7$ and $T_8$: Continuous transitions from quiescent-excitable to active states}{TEXT}}

The transitions represented by $T_7$ and $T_8$ happen between the quiescent excitable state and the active state (only in case C, as illustrated in Fig.~\ref{fig:phase-diagram}c). 
The first occurs through a transcritical-like bifurcation (black dashed lines in Fig.~\ref{fig:phase-diagram}) and the second through a tricritical (or saddle-node-transcritical) point. 
Thus, these two are the counterparts of $T_1$ and $T_2$, respectively, for excitable ---rather than standard--- quiescent states.
 
Let us recall that, as explained above, a naive linearization of the quiescent excitable state (assuming $\Phi>0$) yields eigenvalues with a non-vanishing imaginary part; in any case, the quiescent state remains  stable due to frustrated oscillations that draw the system into the regions II and, possibly, III (Fig.~\ref{fig:regions}b). 

Observe that Eq.\eqref{eq:Transc_E*_delta} and Eq.~\eqref{eq:rho_exd_TDP} remain unchanged for $T_7$ and $T_8$, respectively.
Therefore, the order parameter changes continuously with the control parameter with $\beta = 1$ and $\beta = 1/2$, respectively (Fig.~\ref{fig:beta_exponents}g and \ref{fig:beta_exponents}h).
However, similarly to transition $T_6$, the denominators of Eq.~\eqref{eq:Transc_E*_h} and Eq.~\eqref{eq:rho_exh_TDP}, governing the response to an external field $h$ at the transition point, are negative and the system asymptotically reaches region II, i.e., converges quickly to quiescence. 
Thus, the response to an external field coincides with that of $T_6$, and the exponent $\delta_h$ is not well-defined for either $T_7$ or $T_8$ as there is a discontinuous jump in the order parameter (see Figs.~\ref{fig:beta_exponents}g and~\ref{fig:beta_exponents}h as well as  Figs.~\ref{fig:powerlaws_external_field}g and~\ref{fig:powerlaws_external_field}h).

On the one hand, also as in $T_6$, the asymptotic dynamics of the order parameter in the $T_7$ and $T_8$ transitions exhibit an exponential time decay (Figs.~\ref{fig:powerlaws_time_decay}g and~\ref{fig:powerlaws_time_decay}h). 
On the other hand, the overall behavior of the survival probability, for $T_7$ and $T_8$, shows an intermediate plateau, as opposed to the smooth decay to zero observed for $T_6$ (Figs.~\ref{fig:survival_prob}f-h). 
These plateaus stem from the fact that the excitable quiescent phase is well-established before the transitions take place (in opposition to what happens in $T_6$).

Thus, in summary, the $T_7$ and $T_8$ transitions also exhibit a mixture of features of continuous and discontinuous phase transitions.

\begin{figure*}[!ht]
\centering
{\includegraphics[width = 0.9\textwidth]{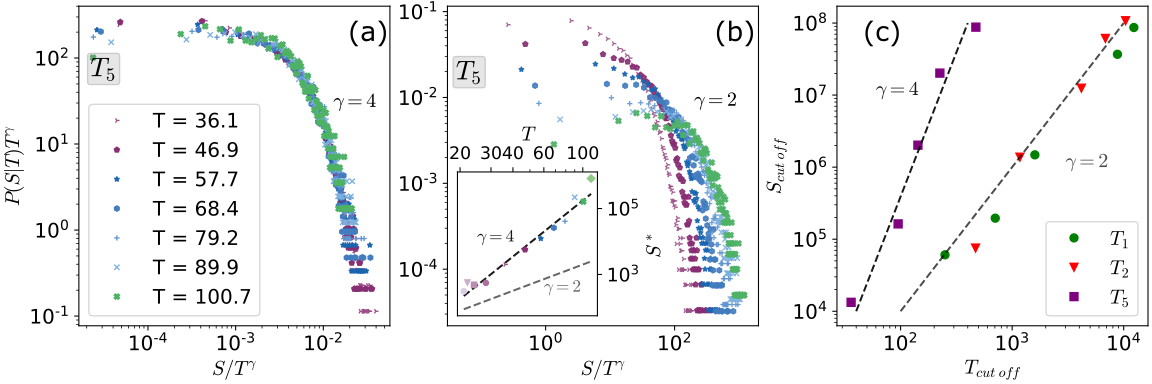}}
\caption{Conditional probability of avalanche sizes given durations, $P(S|T)$, in bonafide continuous phase transitions (i.e., $T_1$, $T_2$, and $T_5$). 
Numerical results obtained with Gillespie's algorithm for the same parameters as Fig.~\ref{fig:phase-diagram}; 
for (a) and (b), the system is poised at $T_5$ with a network size of $N=10^8$; and, for (c), each transition is simulated for sizes $N = 10^4$, $10^5$, $10^6$, $10^7$, and $10^8$.
(a) $P(S|T)$ collapses into a single curve when one sets $\gamma = 4$. 
(b) In contrast, for $\gamma = 2$ there is no curve collapse.
Furthermore, in the inset, one observes that the peaks of $P(S|T)$ ($S^*$) scale with $S^*\sim T^{\gamma}$, with $\gamma = 4$.
(c) Cut-off for size and duration distributions; one scales as a power of the other with the corresponding value of the exponent $\gamma$~\citep{chessa99, DickmanCampelo03}. 
}
\label{fig:prob_cond_inset}
\end{figure*}

\section{The average shape of avalanches}\label{sec:MTP}

The scaling of the mean avalanche shape (also called ``mean temporal profile") of avalanches -- i.e., the fact that the averaged shape of avalanches with different sizes and durations can be collapsed into a single curve by using the adequate value of critical exponents --- has been used as a signature of criticality in non-equilibrium systems with absorbing states~\citep{papanikolaou2011}.
As already mentioned, the DP and TDP universality classes are known to typically have symmetric inverted parabolas mean-temporal profiles of avalanches, a consequence of time-reversal symmetry~\citep{miller2019}.
The asymmetry in the mean temporal profile, when found, reflects a break in such symmetry~\citep{laurson2013}.

In the present Wilson-Cowan model, we observed that, when the transition to quiescence occurs in the neighborhood of $T_5$ (or H+TDP) point in parameter space, avalanche shapes exhibit a non-trivial behavior.
In particular, as illustrated in Fig.~\ref{fig:Skw_tempprofile}a, when one follows the DP ($T_1$) line towards the TDP ($T_2$) point in case A, the mean temporal profile of avalanches acquires only a slight asymmetry (Fig.~\ref{fig:Skw_tempprofile}b), as quantified by the increase in the absolute value of its skewness (Fig.~\ref{fig:Skw_tempprofile}c).
This observation agrees with recent results that show that the introduction of an inhibitory population causes a small tilt on the mean temporal profile at the DP transition~\citep{deCandia21, Corral22, nandi22}.

However,  remarkably, studying the system at the TDP transition as the overall parameters transition from case A to case B (Fig.~\ref{fig:Skw_tempprofile}d), the avalanche mean-temporal-profile becomes progressively more and more asymmetric, with its skewness reaching a maximal absolute value ---i.e., maximal asymmetry as illustrated in Fig.~\ref{fig:Skw_tempprofile}f--- at the H+TDP transition (Fig.~\ref{fig:Skw_tempprofile}e); see also \cite{Corral22}).

\section{Conclusions}

By means of detailed scaling analyses as well as extensive numerical simulations, we have thoroughly analyzed all possible types of  phase transitions between active and quiescent phases that the Wilson-Cowan model exhibits.

On the one hand, under some conditions, the model exhibits the standard phenomenology of systems with quiescent/absorbing states, i.e., two phases (active and quiescent) as well as a phase transition separating them. This transition can be a continuous one (in the mean-field directed-percolation class), a discontinuous one with hysteresis, or a tricritical transition in the tricritical-directed-percolation class, at the point where the previous two types of transitions meet~\citep{Henkel,Odor,Lubeck06}.

On the other hand, a key feature of the mean-field Wilson-Cowan model is that --- in addition to the standard active and quiescent phases --- there is another ``excitable quiescent" phase. 
In particular, the phase diagram describing the model at a mean-field level exhibits a line of Hopf bifurcations to the left of which
there is a standard quiescent state, while, to the
right of it, the convergence towards the quiescent state occurs in an oscillatory way (involving complex eigenvalues). 
Such pseudo-oscillations are nevertheless``frustrated" as the system enters inhibition-dominated regions of the state space (region II or region III in Fig.~\ref{fig:regions}) and, then, activity decays exponentially fast to zero. 
Observe that in this regime, owing to the non-normality of the stability matrix (see below), 
small perturbations to quiescence can be transiently amplified, before decaying back again to quiescence, hence the name ``excitable quiescent" phase (or also, possibly "reactive" phase, see \cite{Benayoun,Serena-NN,Gudowska,Corral22,Hidalgo}).
 
Both of the previous features --- i.e., the presence of a line of Hopf bifurcations and of an excitable-quiescent phase --- stem from the existence of an inhibitory field and cannot possibly appear in simpler models for activity propagation, such as directed percolation or the contact process which include only one field, describing the excitatory activity. 
These two ingredients are at the root of the enriched set of possible phase transitions that the system can exhibit with respect to standard ones.

Our analyses reveal that the Wilson-Cowan model can exhibit three possible types of (bi-dimensional) phase diagrams (as illustrated in Fig.~\ref{fig:phase-diagram}) that can be viewed as sections of a larger (three-dimensional) phase diagram.
These three cases (A, B, and C) differ from one another in the relative position of the (vertical) line of Hopf bifurcations with respect to the tricritical point (and are controlled by a single parameter, $w_{IE}$ in Fig.~\ref{fig:phase-diagram}). 
Careful inspection of the three of them reveals the existence of $8$ different types of phase transitions, labeled $T_1, T_2, \ldots$, and $T_8$, respectively.

Three of them are usual ones separating active from standard quiescent phases and are well-known from the theory of phase transitions:
(i) a (mean-field) directed-percolation (DP), continuous transition ($T_1$); 
(ii) a (mean-field) tricritical directed-percolation (TDP) ($T_2$);
and (iii) a (mean-field) discontinuous transition with bistability and hysteresis ($T_3$).
These three cases correspond to transcritical, saddle-node-transcritical, and saddle-node bifurcations, respectively and exhibit the expected features for their corresponding universality classes.

In particular, let us remark that our results for the DP case ($T_1$) are consistent with those recently reported by de Candia {\it et al.}~\citep{deCandia21}. 
 These authors chose to study the case where $w_{E} \equiv w_{EE} = w_{IE}$ and $w_{I} \equiv w_{EI} = w_{II}$;
 for this choice of parameters, one is in the $T_1$ case (actually, Eq.~\eqref{eq:transc_wEE} becomes $w_{E} - w_{I} = \alpha$, which is the condition for criticality in \citep{deCandia21}).
 Similarly, for $T_2$ our results reproduce the TDP class as first described in~\citep{Lubeck06} (note that recent research has also considered the possibility of a tricritical point in neuronal models with a population of inhibitory units~\citep{Corral22, almeira22}). 
 Finally, for $T_3$, we observe the standard features from  discontinuous phase transitions into absorbing states, such as hysteresis \cite{Lubeck06}. 

Each of the previous three transitions has a counterpart in which the quiescent phase is not a standard one but a quiescent-excitable one: the twin of $T_1$ is a transcritical bifurcation from the quiescent excitable state (labeled $T_7$), the twin of the tricritical $T_2$ is $T_8$, and the twin of $T_3$ is a discontinuous transition, which exhibits bistability between the active and the quiescent-excitable states (labeled $T_4$). 
A peculiar feature of the continuous ones, i.e., $T_7$ and $T_8$, is that their response to an external field is anomalous: even if they are continuous transitions, once the field is introduced they become discontinuous.
In other words, the addition of a small external field drives slightly active states to become quiescent (a phenomenon that stems from the excitability of the quiescent state). 
As a consequence, critical exponents such as $\delta_h$ are not well-defined, so $T_7$ and $T_8$ share features of both continuous and discontinuous phase transitions.

The remaining transitions are unusual and involve entering the active phase right at the point where a Hopf bifurcation also occurs (i.e., they correspond to higher-codimension bifurcations).
In particular, $T_6$ describes the situation in which the Hopf bifurcation falls on top of a transcritical bifurcation, whereas $T_5$ occurs at the special point in which the Hopf bifurcation falls exactly on top of the tricritical point (only possible in case B).

For $T_6$, the transition is adjacent to the excitable-quiescent phase and, thus, one observes the same phenomenon as for $T_7$ and $T_8$, when introducing an external field.
Even if the transition ($T_6$) is continuous, the response to a small external field is anomalous, giving rise to a discontinuity and preventing the exponent $\delta_h$ to be well defined, so again, $T_6$ exhibits features of both continuous and discontinuous phase transitions.

Finally, $T_5$ is by far the most interesting and less trivial transition. 
We have named it Hopf-tricritical directed percolation (H+TDP) transition as it occurs when the Hopf line collides with the tricritical point, giving rise to a codimension 3 transition.
In this case, both eigenvalues of the stability matrix vanish at the transition point, so that the matrix has the normal form of a Bogdanov-Takens bifurcation.
From the point of view of power-counting and dimensionality analyses, this fact has important implications, as carefully discussed above.
In particular, a key aspect is that the scaling features are  controlled by different terms (i) right at criticality and (ii) slightly in the active phase. 
This dichotomy entails a remarkable and surprising violation of some well-established scaling laws. 
It is noteworthy, though, that a breakdown of scaling in a somehow similar model --- including also a second inhibitory-like field --- has been recently reported by
Noh and Park~\citep{NohPark05}.

Another anomalous feature of $T_5$ is that the exponent controlling the growth of the total number of particles in spreading experiments, $\eta$, does not vanish, i.e., $\eta=2$, as opposed to what happens in (mean-field) DP and most other mean-field phase transitions
(as it is related with perturbative corrections to mean-field behavior \cite{Survival1997}).
Moreover, the scaling anomaly of the H+TDP transition is also reflected in its avalanche exponents: while the duration exponent $\tau_t=2$ is consistent with DP and TDP, the size-distribution exponent $\tau=5/4$ and crackling noise exponent $\gamma = 4$ are different from the usual ones ($\tau=3/2$ and $\gamma = 2$, respectively).

A more systematic field-theoretical analysis of the H+TDP universality class --- as well as its implementation in finite-dimensional substrates --- is left for future work.

A relevant hallmark of standard models in the DP class is the symmetry in the mean temporal profile of avalanches, which reveals time-reversal invariance~\citep{laurson2013, miller2019}. 
On the contrary, the mean temporal profile in $T_5$ shows a strong asymmetry that we have quantified in terms of negative skewness. 
Previous work has shown that the introduction of inhibition tilts the once symmetric parabolas produced by models in the DP universality class~\citep{Corral22, nandi22}.
We further propose that not only the strength of the inhibitory  coupling slightly tilts the mean temporal profiles, but that the combination of the proximity to the excitable quiescent phase and to the onset of frustrated oscillations promotes even greater distortions. 
Considering the inherent difficulties in assessing avalanche exponents from experimental data, the asymmetry in the avalanche shape collapse may turn out to be a useful additional tool to more directly reveal proximity to this anomalous transition.

Last but not least, it is also worth stressing that the nature of the phase diagram and phase transitions that we have reported for the mean-field Wilson-Cowan model may change when sparse networks are considered~\citep{Buendia-Jensen}. 
In particular, the presence of enhanced stochastic effects, stemming from the finite connectivity of each unit, can significantly alter the dynamics and induce novel phenomena~\cite{Corral22}. 
The  study of the interplay between the transitions discussed here and such additional stochastic effects remains to be pursued.
Similarly, the effect of structural
heterogeneity, e.g. the presence of local excitation/inhibition imbalances, that could potentially lead to extended critical (Griffiths) phases \cite{GPCN,Moretti,Odor-GP}, remains as an open  challenge for future work.

\begin{acknowledgments}

\vspace{1.0cm}
MAM acknowledges the Spanish Ministry and Agencia Estatal de investigaci{\'o}n (AEI) through Project of I+D+i Ref. PID2020-113681GB-I00, financed by
MICIN/AEI/10.13039/501100011033 and FEDER “A way to make Europe”, as well as the Consejer{\'\i}a de Conocimiento, Investigaci{\'o}n Universidad, Junta de Andaluc{\'\i}a and European Regional Development Fund, Project P20-00173 for financial support.
HCP acknowledges CAPES (PrInT grant 88887.581360/2020-00) and thanks the hospitality of the Statistical Physics group at the Instituto Interuniversitario Carlos I de F\'isica Te\'orica y Computacional at the University of Granada during her six-month stay, during which part of this work was developed.
MC acknowledges support by CNPq (grants 425329/2018-6 and  301744/2018-1), CAPES (grant PROEX 23038.003069/2022-87), and FACEPE (grant APQ-0642-1.05/18).
This article was produced as part of the activities of Programa Institucional de Internacionaliza\c{c}\~ao (PrInt).
We are also very thankful to R. Corral, S. di Santo, V. Buendia, J. Pretel, and I. L. D. Pinto for valuable discussions and comments on  previous versions of the manuscript.
\end{acknowledgments}
\vspace{2cm} 

\subsection{Appendix: Gillespie's algorithm}

The stochastic version of the Wilson-Cowan model \cite{Benayoun} was simulated using Gillespie's algorithm~\citep{GILLESPIE76, Gillespie77}, following these steps:
\begin{description}
\item[Step 0] initialize the system; for spreading experiments and avalanche analyses, only an excitatory site is active at $t = 0$;
\item[Step 1] at each time step, calculate the transition rates for each neuron --- if active, $\Phi(s_i)$, and otherwise, $\alpha$ --- and add these rates, $r = \sum_i r_i$;
\item[Step 2] the time step is chosen from an exponential distribution with rate $r$ and added to the total-time counter;
\item[Step 3] and, the site to be updated is chosen with probability $r_i/r$, where $r_i$ is the transition rate of the neuron.
\end{description}
The size (duration) of an avalanche is counted as the total number of activations (total time) of a single instance of the simulation starting from just one excitatory activated site before returning to quiescence.

\subsection{Appendix: Mathematical conditions for the bifurcation lines/points}\label{ap:mat_TDP}

The mathematical condition for the tricritical point is derived from a standard linear-stability analysis of the stationary solution around zero. 
First of all, observe that Eq.~\eqref{eq:EI_fixed_point} and Eq.~\eqref{eq:IE_fixed_point} have positive solutions. 
One can express $w_{EE}$ as a function of the fixed-point solution $(E^*, I^*)$ as specified by Eq.~\eqref{eq:w_EE_function}:
\begin{eqnarray}
w_{EE} &=& \frac{1}{E^*}\left[w_{EI}I^* + \Phi^{-1}\left(\frac{\alpha E^*}{1 - E^*}\right)\right].
\end{eqnarray}
Expanding in power series around the origin one can readily verify the emergence of an active-state solution at the value of $w^T_{EE}$, specified in Eq.~\eqref{eq:transc_wEE}.
For values $w_{EE} > w^{T}_{EE}$, the origin loses stability to a positive solution in a transcritical bifurcation.
Defining the distance to the critical value of the control parameter, $\Delta = w_{EE} - w_{EE}^T$, the value of this non-trivial solution scales linearly with $\Delta$ as 
\begin{eqnarray}
E^* \sim
I^* \sim [(\alpha + w_{II})^3 - w_{EI}w_{IE}^2]^{-1}\Delta.
\label{eq:E_delta_transc}
\end{eqnarray}
Since $\Delta \geq 0$, this solution is positive for $(\alpha + w_{II})^3 - w_{EI}w_{IE}^2 > 0$.

Observe that, for $(\alpha + w_{II})^3 - w_{EI}w_{IE}^2 = 0$, Eq.~\eqref{eq:E_delta_transc} diverges.
The saddle-node and transcritical bifurcations collide into a saddle-node transcritical (SNT) bifurcation or tricritical point~\citep{SNT2019,SNT2020}.
Observe that in Fig.~\ref{fig:phase-diagram}, a black circle marks the tricritical point in all cases (i.e.,  $T_2$, $T_5$, and $T_8$).
In cases A and C, this bifurcation has codimension 2 and occurs at:
\begin{eqnarray}
w^{SNT}_{EI} &=& \frac{(\alpha + w_{II})^3}{w_{IE}^2},
\label{eq:SNT_wei_A}\\
w^{SNT}_{EE} &=& \alpha + \frac{(\alpha + w_{II})^2}{w_{IE}}\; .
\label{eq:SNT_wee_A}
\end{eqnarray}
The non-trivial solution emerges from the trivial solution with $w_{EE}$ and it scales with the distance to the critical value, $\Delta$, as $E^* \propto I^* \propto \Delta^{1/2}$.

Finally, in Fig.~\ref{fig:phase-diagram} case B, a codimension 3 bifurcation emerges from an extra fine-tuning of the parameters when $w_{IE} = \alpha + w_{II}$.
For this choice of parameters, at $w_{EI} = w_{IE}$, the saddle-node transcritical collides with the Hopf right at the tricritical point, $T_5$.
Combining Eq.~\eqref{eq:AHT_wei} and Eq.~\eqref{eq:SNT_wei_A}, the values of the control parameters, for this bifurcation, are:
\begin{eqnarray}
w_{EI} &=& \alpha + w_{II}\label{SNTAH_wchi}\\
w_{EE} &=& 2\alpha + w_{II}\label{SNTAH_wee}.
\end{eqnarray}

\subsection{Appendix: Do trajectories \emph{cross or slide} onto the switching manifolds?}\label{ap:trap}

Piece-wise continuous dynamics have two possible behaviors at the switching manifolds: sliding or crossing~\citep{Piecewise19}.
To determine the behavior of the Wilson-Cowan model system, we consider the Heaviside function, Eq.~\eqref{eq:response_function}, in Eq.~\eqref{eq:WCEI-Eeq} and Eq.~\eqref{eq:WCEI-Ieq}:
\begin{eqnarray}
\dot{x} = \left\{
\begin{array}{l}
f^+_x \equiv -\alpha x + (1 - x)\tanh(w_ix - w_jy)\textrm{ ,}  \\
\hspace{.8in}\textrm{if } s \equiv w_ix - w_jy > 0\\
f^-_x \equiv -\alpha x\textrm{ ,}\hspace{.1in} \textrm{if }s < 0
\end{array}
\right. ,
\end{eqnarray}
where $f^+_x$ ($f^-_x$) is evaluated to the right (left) of the switching manifold, $s = 0$.

Let us consider the switching manifold $s_{E} = 0$, where $E = (w_{EI} / w_{EE})I $, Fig.~\ref{fig:regions}b. 
One can then write:
\begin{eqnarray}
\vec\nabla s_E &=& \left(
\begin{array}{c}
\frac{\partial}{\partial E}s_E \\
\frac{\partial}{\partial I}s_I
\end{array}
\right) = \left(
\begin{array}{c}
W_{EE} \\
-W_{EI}
\end{array}
\right)\\
\vec{f}^+&=& \left(
\begin{array}{c}
-\alpha E + (1 - E)
\tanh\left(w_{EE}E - w_{EI}I\right) \\
-\alpha I + (1 - I)
\tanh\left(w_{IE}E - w_{II}I\right)
\end{array}
\right)^T\\
\vec{f}^-&=& \left(
\begin{array}{c}
-\alpha E\\
-\alpha I + (1 - I)
\tanh\left(w_{IE}E - w_{II}I\right)
\end{array}
\right)^T\; .
\end{eqnarray}
In order to know if when the system reaches the switching manifold the trajectories will cross it or slide on it one needs to evaluate the sign of $(\vec{f}^+ \cdot \vec\nabla s_E)(\vec{f}^- \cdot \vec\nabla s_E)$ at the switching manifold 
\begin{eqnarray}
\vec{f}^+ \cdot \vec\nabla s_E &=& -\alpha(w_{EE}E - w_{EI}I) +\nonumber\\
&&+ w_{EE}(1 - E)\tanh\left(w_{EE}E - w_{EI}I\right) +\nonumber\\
&&- w_{EI}(1 - I)\tanh\left(w_{IE}E - w_{II}I\right)\\
\vec{f}^- \cdot \vec\nabla s_E &=& -\alpha(w_{EE}E - w_{EI}I) +\nonumber\\
&&- w_{EI}(1 - I)\tanh\left(w_{IE}E - w_{II}I\right)
\end{eqnarray}
Given that at the switching manifold, $w_{EE}E = w_{EI}I$:
{\medmuskip=0mu
\thinmuskip=0mu
\thickmuskip=0mu
\begin{eqnarray}
\vec{f}^+ \cdot \vec\nabla s_E &=& - w_{EI}(1 - I)\times\nonumber\\
&&
\tanh\left(\frac{w_{IE}(w_{EI} - w_{II})}{w_{EE}}I\right)\\
\vec{f}^- \cdot \vec\nabla s_E &=&- w_{EI}(1 - I)\times\nonumber\\
&&
\tanh\left(\frac{w_{IE}(w_{EI} - w_{II})}{W_{EE}}I\right),
\end{eqnarray}
\begin{eqnarray}
(\vec{f}^+ \cdot \vec\nabla s_E)(\vec{f}^- \cdot \vec\nabla s_E) &&= \left[w_{EI}(1 - I)\times\right.\nonumber\\
&&\left.
\tanh\left(\frac{w_{IE}(w_{EI} - w_{II})}{w_{EE}}I\right)\right]^2.
\end{eqnarray}
}
For $(\vec{f}^+ \cdot \vec\nabla s_E)(\vec{f}^- \cdot \vec\nabla s_E) > 0$, the trajectories cross the switching manifold, creating a trapping region.

\bibliography{references_}

\end{document}